\documentclass[a4paper,11pt,numbers=noenddot]{scrartcl}
\addtokomafont{disposition}{\rmfamily}
\usepackage[utf8x]{inputenc}      %
\usepackage{amsmath,amssymb}
\usepackage[T1]{fontenc}          %
\usepackage{booktabs,tabularx}
\usepackage{ifthen}
\usepackage{multirow}
\usepackage{graphicx}
\usepackage{scalefnt}
\usepackage{xspace}
\usepackage{listings}
\usepackage[many]{tcolorbox}
\usepackage[numbers,sort&compress]{natbib} %
\usepackage[font=small,labelfont=bf,format=plain,margin=0.05\textwidth]{caption}
\usepackage[pdftitle={Higgs mass prediction in the MSSM at three-loop
    level in a pure DRbar context},
  pdfauthor={Robert V. Harlander, Jonas Klappert, Alexander Voigt},
  pdfkeywords={H3m,FlexibleSUSY,Higgs,supersymmetry},
  bookmarks=true, linktocpage,
  colorlinks=true, allbordercolors=white, allcolors=blue]{hyperref}

\newcommand{\EFT}{\abbrev{EFT}\@\xspace}
\newcommand{\citere}[1]{Ref.\,\cite{#1}}
\newcommand{\citeres}[1]{Refs.\,\cite{#1}}
\newcommand{\eqn}[1]{Eq.\,\eqref{#1}}
\newcommand{\eqns}[1]{Eqs.\,\eqref{#1}}
\newcommand{\drbarmass}{m}

\newcommand{\mstop}[1]{\drbarmass_{\tilde t\ifthenelse{\equal{#1}{}}{}{, }#1}}
\newcommand{\mstopmod}[1]{\hat{\drbarmass}_{\tilde t\ifthenelse{\equal{#1}{}}{}{, }#1}}
\newcommand{\msbotmod}[1]{\hat{\drbarmass}_{\tilde b\ifthenelse{\equal{#1}{}}{}{, }#1}}
\newcommand{\msquark}[1][]{\drbarmass_{\tilde q\ifthenelse{\equal{#1}{}}{}{, }#1}}
\newcommand{\one}{one}
\newcommand{\two}{two}
\newcommand{\three}{three}
\newcommand{\abbrev}[1]{{\scalefont{.9}\text{#1}}}
\newcommand{\himalaya}{\texttt{Himalaya}\@\xspace}
\newcommand{\htm}{\texttt{H3m}\@\xspace}
\newcommand{\fs}{\texttt{Flex\-i\-ble\-SUSY}\@\xspace}
\newcommand{\fsh}{\texttt{FlexibleSUSY+\allowbreak\himalaya}\@\xspace}
\newcommand{\HSSUSY}{\texttt{HSSUSY}\@\xspace}
\newcommand{\sarah}{\texttt{SARAH}\@\xspace}
\newcommand{\spheno}{\texttt{SPheno}\@\xspace}
\newcommand{\susyhd}{\texttt{SusyHD}\@\xspace}
\newcommand{\MhEFT}{\texttt{MhEFT}\@\xspace}
\newcommand{\softsusy}{\texttt{SOFTSUSY}\@\xspace}

\newcommand{\FeynHiggs}{\texttt{FeynHiggs}\@\xspace}

\newcommand{\CPsuperH}{\texttt{CPsuperH}\@\xspace}
\newcommand{\SuSpect}{\texttt{SuSpect}\@\xspace}
\newcommand{\ISASUSY}{\texttt{ISASUSY}\@\xspace}

\newcommand{\mathematica}{\texttt{Mathematica}\xspace}
\newcommand{\code}[1]{\lstinline|#1|}  %
\newcommand{\ol}[1]{\overline{#1}}
\newcommand{\MSbar}{\ensuremath{\ol{\abbrev{MS}}}\xspace}
\newcommand{\DRbar}{\ensuremath{\ol{\abbrev{DR}}}\xspace}
\newcommand{\MDRbar}{\ensuremath{\ol{\abbrev{MDR}}}\xspace}
\newcommand{\unit}[1]{\,\text{#1}}      %

\newcommand{\SM}{\ensuremath{\abbrev{SM}}\xspace}

\newcommand{\SUSY}{\ensuremath{\abbrev{SUSY}}\xspace}
\newcommand{\QCD}{\ensuremath{\abbrev{QCD}}\xspace}
\newcommand{\SQCD}{\ensuremath{\abbrev{SQCD}}\xspace}
\newcommand{\MSSM}{\ensuremath{\abbrev{MSSM}}\xspace}
\newcommand{\LHC}{\ensuremath{\abbrev{LHC}}\xspace}
\newcommand{\susy}{\SUSY}
\newcommand{\mssm}{\MSSM}
\newcommand{\sm}{\SM}

\newcommand{\MS}{\ensuremath{M_S}}

\newcommand{\QED}{\ensuremath{\abbrev{QED}}\xspace}
\newcommand{\tree}{\text{tree}}

\newcommand{\figref}[1]{\figurename~\ref{#1}}
\newcommand{\figsref}[1]{\figurename s~\ref{#1}}
\newcommand{\secref}[1]{Section~\ref{#1}}
\newcommand{\tabref}[1]{\tablename~\ref{#1}}
\newcommand{\ptitle}[1]{\emph{#1}}
\renewcommand{\ptitle}[1]{}
\newcommand{\Fortran}{\texttt{FORTRAN}\xspace}

\newcommand{\order}[1]{\ensuremath{\mathcal{O}\!\left(#1\right)}}

\makeatletter
\lstnewenvironment{numlstlisting}[1][]{%
  \lstset{%
    #1,
    numbers=left,
    firstnumber=auto,
    numberstyle=\tiny\sffamily}%
  \csname\@lst @SetFirstNumber\endcsname
}{%
  \csname \@lst @SaveFirstNumber\endcsname
}
\makeatother

\DeclareMathOperator{\re}{Re}

\def\at{\alpha_t}
\def\ab{\alpha_b}
\def\as{\alpha_s}
\def\atau{\alpha_{\tau}}
\def\aem{\alpha_{\text{em}}}
\def\atas{\at\as}
\def\asas{\as^2}
\def\atasas{\at\asas}
\def\abasas{\ab\asas}
\def\oat{\order{\at}}

\def\oatas{\order{\at\as}}
\def\oabas{\order{\ab\as}}

\begin{document}

\thispagestyle{empty}
\begin{flushleft}
TTK--17--25\\
August 2017
\end{flushleft}
\vskip 2 cm
\begin{center}
  {\Large \textbf{Higgs mass prediction in the MSSM
      at \three-loop level
    in a pure \DRbar context \\[.3em]}}
\bigskip
\bigskip
{
Robert V. Harlander, Jonas Klappert, Alexander Voigt
}
\bigskip
\vspace{0.23cm}\\
{\textit{Institute for Theoretical Particle Physics and Cosmology, RWTH
    Aachen University, 52074 Aachen, Germany}\\
\bigskip
\vspace{0.23cm}
\texttt{harlander@physik.rwth-aachen.de}\\
\texttt{klappert@physik.rwth-aachen.de}\\
\texttt{avoigt@physik.rwth-aachen.de}}
\bigskip
\end{center}

\begin{abstract}
  The impact of the \three-loop effects of order $\at\as^2$ on the mass
  of the light \abbrev{CP}-even Higgs boson in the \MSSM is studied in a
  pure \DRbar context. For this purpose, we implement the results of
  Kant\,\textit{et al.}\,\cite{Kant:2010tf} into the \texttt{C++} module
  \himalaya and link it to \fs, a \mathematica and \texttt{C++} package
  to create spectrum generators for \abbrev{BSM} models.  The
  \three-loop result is compared to the fixed-order \two-loop
  calculations of the original \fs and of \FeynHiggs, as well as to the
  result based on an \EFT approach.  Aside from the expected reduction
  of the renormalization scale dependence with respect to the lower
  order results, we find that the \three-loop contributions
  significantly reduce the difference from the \EFT prediction in the
  TeV-region of the \SUSY scale $\MS$.  \himalaya can be linked also to
  other \two-loop \DRbar codes, thus allowing for the elevation of these
  codes to the \three-loop level.
\end{abstract}

\clearpage
\section{Introduction}\label{sec:intro}

The measurement of the Higgs boson mass at the Large Hadron Collider
(\LHC) represents a significant constraint on the viability of
supersymmetric (\SUSY) models.  Given a particular \SUSY model, the mass
of the Standard Model-like Higgs boson is a prediction, which must be in
agreement with the measured value of $(125.09 \pm 0.21
\pm 0.11)\unit{GeV}$ \cite{Aad:2015zhl}.  Noteworthy, the experimental
uncertainty on the measured Higgs mass has already reached the per-mille
level.  Theory predictions in \SUSY models, however, struggle to reach
the same level of accuracy.  The reason is that the Higgs mass receives
large higher order corrections, dominated by the top Yukawa and the
strong gauge
coupling\,\cite{Haber:1990aw,Ellis:1990nz,Heinemeyer:1998jw}.  Both of
these two couplings are comparatively large, leading to a relatively
slow convergence of the perturbative series.  Furthermore, the scalar
nature of the Higgs implies corrections proportional to the square of
the top-quark mass, on top of the top-mass dependence due to the Yukawa
coupling, which enters the loop corrections quadratically.  On the other
hand, corrections from \susy{} particles are only logarithmic in
the \susy{} particle masses due to the assumption of only
soft \SUSY-breaking terms.  If the \SUSY particles are not too far above
the TeV scale \cite{ATLAS:2017kyf,ATLAS:2017tpg}, the \susy Higgs mass
can be obtained from a fixed-order calculation of the relevant one- and
two-point functions with external Higgs fields. In this case, higher
order corrections up to the \three-loop level are known in the Minimal
Supersymmetric Standard Model (\mssm)
\cite{Heinemeyer:1998jw,
  Heinemeyer:1998kz,Heinemeyer:1998np,Zhang:1998bm,Espinosa:1999zm,
  Degrassi:2001yf,Brignole:2001jy,Dedes:2002dy,
  Brignole:2002bz,Dedes:2003km,Espinosa:2000df,Heinemeyer:2004xw,
  Martin:2002wn,Harlander:2008ju,Kant:2010tf,
  Borowka:2014wla,Borowka:2015ura,Degrassi:2014pfa}.

There are plenty of publicly available computer codes which calculate the
Higgs pole mass(es) in the \MSSM at higher orders:
\CPsuperH\ \cite{Lee:2003nta,Lee:2007gn,Lee:2012wa}, \FeynHiggs\
\cite{Heinemeyer:1998yj,Heinemeyer:1998np,Degrassi:2002fi,Frank:2006yh,Hahn:2013ria,Bahl:2016brp},
\fs\ \cite{Athron:2014yba,Athron:2016fuq}, \htm\
\cite{Harlander:2008ju,Kant:2010tf}, \ISASUSY\ \cite{Baer:1993ae},
\MhEFT\ \cite{Lee:2015uza}, \sarah/\spheno\
\cite{Porod:2003um,Staub:2009bi,
  Porod:2011nf,Staub:2010jh,Staub:2012pb,
  Staub:2013tta,Staub:2017jnp}, \softsusy\
\cite{Allanach:2001kg,Allanach:2014nba}, \SuSpect\
\cite{Djouadi:2002ze} and \susyhd\ \cite{Vega:2015fna}.
%
\FeynHiggs adopts the on-shell scheme for the renormalization of the
particle masses, while all other codes express their results in terms
of \MSbar/\DRbar parameters.  All these schemes are formally
equivalent up to higher orders in perturbation theory, of course. The
numerical difference between the schemes is one of the sources of
theoretical uncertainty on the Higgs mass prediction, however.
All of these programs take into account \one-loop corrections, most of
them also leading \two-loop corrections. \htm is the only one which includes
\three-loop corrections of order $\atasas$, where $\at$ is the squared
top Yukawa and $\as$ is the strong coupling. It combines these terms
with the on-shell \two-loop result of \FeynHiggs after transforming the
$\oat$ and $\oatas$ terms from there to the \DRbar scheme.

Here we present an alternative implementation of the $\order{\atasas}$
contributions of \citeres{Harlander:2008ju,Kant:2010tf} for the light
\abbrev{CP}-even Higgs mass in the \MSSM into the framework of
\fs\,\cite{Athron:2014yba}, referring to the combination as \fsh in what
follows. This allows us to study the effect of the \three-loop
contributions in a pure \DRbar environment, i.e.\ without the trouble of
combining the corrections with an on-shell calculation.  The \three-loop
terms are provided in the form of a separate \texttt{C++} package, named
\himalaya, which one should be able to include in any other \DRbar code without
much effort.  The \himalaya package and the dedicated version of \fs
which
incorporates the \three-loop contributions from \himalaya, can be
downloaded from \citeres{himalaya,flexiblesusy},
respectively. In this way, we hope to contribute to the on-going effort
of improving the precision of the Higgs mass prediction in the \MSSM.

In the present paper we study the impact of the \three-loop corrections
for low and high \SUSY scales and compare our results to the \two-loop
calculations of the public spectrum generators of \fs and \FeynHiggs.
By quantifying the size of the \three-loop corrections, we also provide
a measure for the theoretical uncertainty of the \DRbar fixed-order
calculation.

As will be shown below, the implementation of the $\atasas$ corrections
also applies to the terms of order $\abasas$, where $\ab$ is the bottom
Yukawa coupling. Therefore, \himalaya will take such terms into account,
and we will refer to the sum of top- and bottom-Yukawa induced
supersymmetric \QCD (\SQCD) corrections as $\order{\atasas+\abasas}$ in
what follows. However, it should be kept in mind that this does not
include effects of order $\as^2\sqrt{\at\ab}$, which arise from
\three-loop Higgs self energies involving both a top/stop and a
bottom/sbottom triangle. The results of \himalaya are thus unreliable in
the (rather exotic) case where $\at$ and $\ab$ are comparable in
magnitude.

The remainder of this paper is structured as
follows. Section\,\ref{sec:higgsMass3L} describes the form in which
the \three-loop contributions of order $(\at+\ab)\as^2$ are implemented
in \himalaya. Its input parameters are to be provided in the \DRbar
scheme at the appropriate perturbative order. Section\,\ref{sec:fs}
details how this input is prepared in the framework of \fs. It also
summarizes all the contributions that enter the final Higgs mass
prediction in \fsh. Section\,\ref{sec:results} analyzes the impact of
various \three-loop contributions on this prediction as well as the
residual renormalization scale dependence, and it compares the results
obtained with \fsh to existing fixed-order and resummed results for the
light Higgs mass. In particular, this includes a comparison to the
original implementation of the \three-loop effects in \htm. Our
conclusions are presented in Section\,\ref{sec:conclusions}. Technical
details of \himalaya, its link to \fs, and run options are collected in
the appendix.

\section{Higgs mass prediction at the \three-loop level
in the \MSSM}\label{sec:higgsMass3L}

The results for the \three-loop $\atasas$ corrections to the Higgs mass
in the \MSSM\ have been obtained in
\citeres{Harlander:2008ju,Kant:2010tf} by a Feynman diagrammatic
calculation of the relevant one- and two-point functions with external
Higgs fields in the limit of vanishing external momenta. The dependence
of these terms on the squark and gluino masses was approximated through
asymptotic expansions, assuming various hierarchies among the masses of
the \susy\ particles.  For details of the calculation we refer to
\citeres{Harlander:2008ju,Kant:2010tf}.

\subsection{Selection of the hierarchy}\label{sec:hierarchy}

A particular set of parameters typically matches several of the
hierarchies mentioned above. In order to select the most suitable one,
\citere{Kant:2010tf} suggested a pragmatic approach, namely the
comparison of the various asymptotic expansions to the exact expression
at two-loop level. \himalaya also adopts this approach, but introduces a
few refinements in order to further stabilize the hierarchy selection
(see also \citere{Pak:2012xr}).

In a first step the Higgs pole mass $M_h$ is calculated at the two-loop
level at order $\atas$ using the result of
\citere{Degrassi:2001yf} in the form of the associated
\Fortran code provided by the authors. We refer to this quantity
as $M_h^\text{DSZ}$ in what follows. Subsequently, for all hierarchies
$i$ which fit the given mass spectrum, $M_h$ is calculated again using
the expanded expressions of \citere{Kant:2010tf} at the two-loop level,
resulting in $M_{h,i}$. In the original approach of
\citere{Kant:2010tf}, the hierarchy is selected as the value of $i$
for which the difference
\begin{equation}
\delta^{\mathrm{2L}}_i = \left|M^\text{DSZ}_{h} - M_{h,i}\right|
\label{eq:dsz_error}
\end{equation}
is minimal. However, we found that this criterion alone causes
instabilities in the hierarchy selection in regions where several
hierarchies lead to similar values of $\delta^\mathrm{2L}_i$. We
therefore refine the selection criterion by taking into account the
quality of the convergence in the respective hierarchies, quantified by
\begin{equation}
\delta^{\mathrm{conv}}_i = \sqrt{\sum_{j=1}^n\left(M_{h,i} - M^{(j)}_{h,i}\right)^2}\,.
\label{eq:exp_error}
\end{equation}
While $M_{h,i}$ includes all available terms of the expansion in mass
(and mass difference) ratios, in $M^{(j)}_{h}$ the highest terms of the
expansion for the mass (and mass difference) ratio $j$ are dropped.  We
then define the ``best'' hierarchy to be the one which minimizes the
quadratic mean of \eqns{eq:dsz_error} and \eqref{eq:exp_error},
\begin{equation}
\delta_i = \sqrt{\left(\delta_i^{\mathrm{2L}}\right)^2 +
  \left(\delta_i^{\mathrm{conv}}\right)^2}\,.
\label{eq:deltai-definition}
\end{equation}
The relevant analytical expressions for the three-loop terms of order
$\at\as^2$ to the
\abbrev{CP}-even Higgs mass matrix in the various mass hierarchies are
quite lengthy. However, they are accessible in \mathematica
format in the framework of the publicly available
program \htm. We have transformed these formulas into \texttt{C++}
format and implemented them into \himalaya.

The hierarchies defined in \htm equally apply to the top and the bottom
sector of the \mssm, so that the results of \citere{Kant:2010tf} can
also be used to evaluate the corrections of order $\ab\as^2$ to the
Higgs mass matrix. Indeed, \himalaya takes these corrections into
account. However, as already pointed out in Section\,\ref{sec:intro}, a
complete account of the top- \textit{and} bottom-Yukawa effects to order
$\as^2$ would require to include the contribution of diagrams which
involve both top/stop and bottom/sbottom loops at the same time. These
were not considered in \citere{Kant:2010tf}, and thus the \himalaya
result should only be used in cases where such mixed $\sqrt{\at\ab}$
terms can be neglected.

\subsection{Modified $\DRbar$ scheme}
\label{sec:MDR_scheme}

By default, all the parameters of the calculation are renormalized in
the \DRbar\ scheme. However, in this scheme, one finds artificial
``non-decoupling'' effects\,\cite{Degrassi:2001yf}, meaning that the
\two- and \three-loop result for the Higgs mass depends quadratically on
a \susy particle mass if this mass gets much larger than the others.
Such terms are avoided by transforming the stop masses to a non-minimal
scheme, named $\MDRbar$ (modified $\DRbar$) in \citere{Kant:2010tf},
which mimics the virtue of the on-shell scheme of automatically
decoupling the heavy particles.

If the user wishes to use this scheme rather than pure \DRbar, \himalaya
writes the Higgs mass matrix as
\begin{equation}
  \begin{split}
    \hat{\mathsf{M}}(\mstopmod{}) &=
    \hat{\mathsf{M}}^\tree
    + \hat{\mathsf{M}}^{(\at)}(\mstopmod{})
    + \hat{\mathsf{M}}^{(\atas)}(\mstopmod{})
    + \hat{\mathsf{M}}^{(\atasas)}(\mstopmod{}) + \cdots\\
    &=
    \mathsf{M}^\tree
    + \mathsf{M}^{(\at)}(\mstop{})
    + \mathsf{M}^{(\atas)}(\mstop{})
    + \delta\mathsf{M}(\mstop{},\mstopmod{})
    + \hat{\mathsf{M}}^{(\atasas)}(\mstopmod{}) + \cdots ,
    \label{eq:Mmdr}
  \end{split}
\end{equation}
where $\mathsf{M}$ and $\hat{\mathsf{M}}$ are the Higgs mass matrices in
the \DRbar and the \MDRbar scheme, respectively,
$\mathsf{M}^\text{tree}=\hat{\mathsf{M}}^\text{tree}$ is the tree-level
expression, and the superscript ${}^{(x)}$ denotes the term of order
$x\in\{\at,\as,\atas,\ldots\}$. The ellipsis in \eqn{eq:Mmdr} symbolizes
any terms that involve coupling constants other than $\at$ or $\as$, or
higher orders of the latter.  For brevity we suppress the stop mass
indices ``1'' and ``2'' here.
\himalaya provides the numerical results for
$\hat{\textsf{M}}^{(\atasas)}(\mstopmod{})$ as well as
\begin{equation}
  \begin{split}
    \delta\mathsf{M}(\mstop{},\mstopmod{}) \equiv
    \left(\hat{\mathsf{M}}^{(\at)}(\mstopmod{})
      + \hat{\mathsf{M}}^{(\atas)}(\mstopmod{})\right)
    -\left(\mathsf{M}^{(\at)}(\mstop{})
      +\mathsf{M}^{(\atas)}(\mstop{})\right)\,,
  \end{split}
\label{eq:DRToMDRShift}
\end{equation}
where the \MDRbar stop mass $\mstopmod{}$ is calculated from its \DRbar
value $\mstop{}$ by the conversion formulas through $\order{\as^2}$,
provided in
\citere{Kant:2010tf}. Note that these conversion formulas depend on
the underlying hierarchy, and may be different for $\mstop{1}$ and
$\mstop{2}$.

Even if the result is requested in the \MDRbar scheme, the output
of \himalaya can thus be directly combined with pure \DRbar results
through $\order{\atas}$ according to \eqn{eq:Mmdr} in order to arrive at
the mass matrix at order $\atasas$. Of course, one may also request the
plain \DRbar result from \himalaya, in which case it will simply return
the numerical value for $\mathsf{M}^{(\atasas)}(\mstop{})$ which can
be directly added to any \two-loop \DRbar result.

\begin{figure}[tbh]
  \centering
  \includegraphics[width=0.49\textwidth]{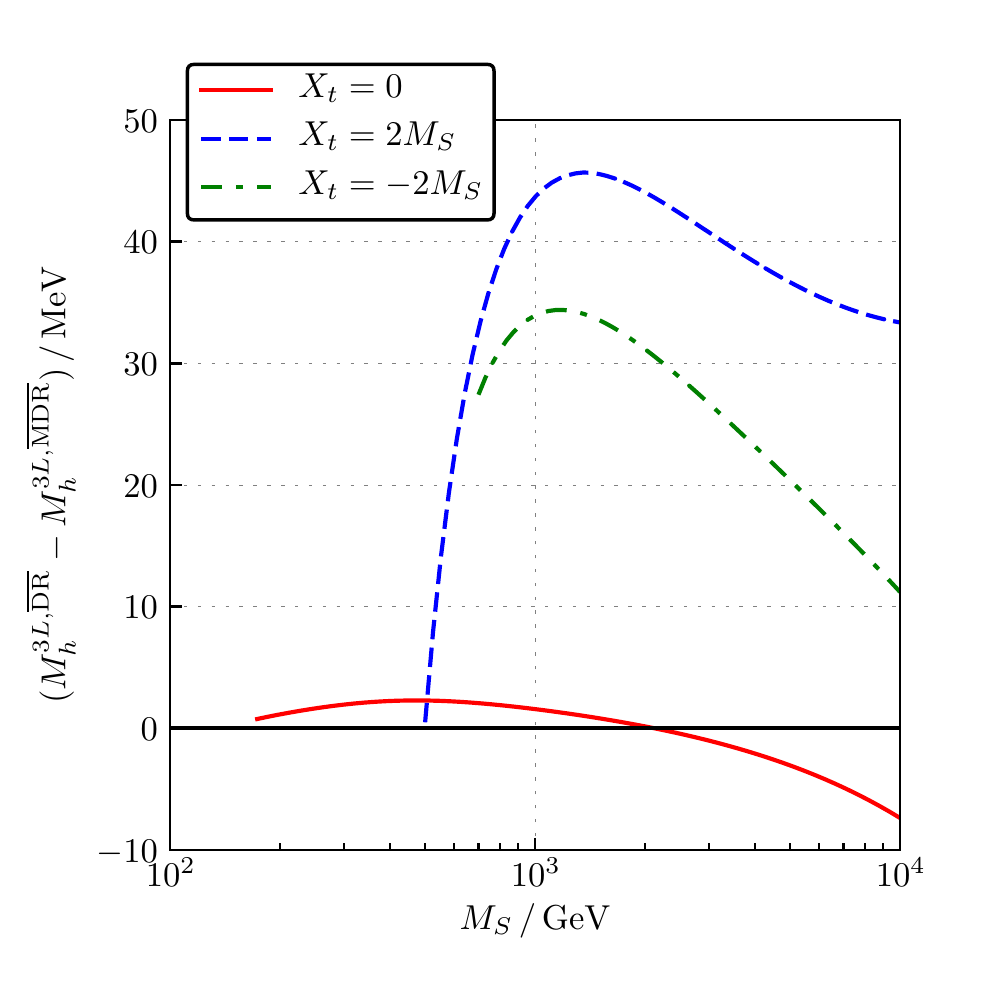}\hfill
  \includegraphics[width=0.49\textwidth]{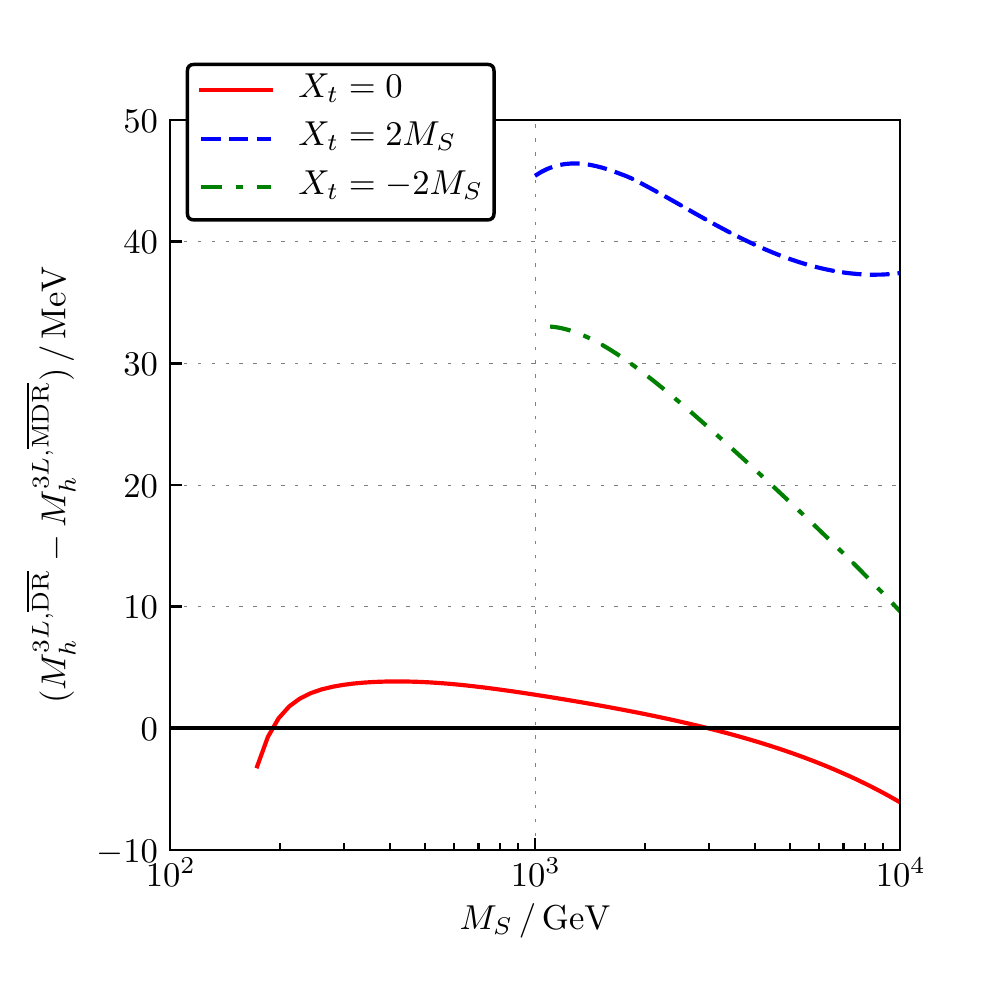}
  \caption{Difference between the lightest Higgs pole mass
    calculated in the \DRbar scheme and the \MDRbar scheme as a
    function of the \SUSY scale $\MS$ for $\tan\beta = 5$.  In the left panel the
    soft-breaking stop and gluino mass parameters are set equal to
    $\MS$.  In the right panel, we use $m_{\tilde{g}} = 2\MS$.  We
    have cut off curves with non-zero $X_t$ around or below the TeV
    scale, where the \DRbar\ \abbrev{CP}-even Higgs mass becomes
    tachyonic at the electroweak scale.}
  \label{fig:DR_vs_MDR}
\end{figure}

In any case, the difference between the \DRbar and \MDRbar result is
expected to be quite small unless the mass splitting between one of the
stop masses and other, heavier, strongly interacting \SUSY particles
becomes very large. As a practical example, in \figref{fig:DR_vs_MDR} we
show the difference of the lightest Higgs mass at the \three-loop level
calculated in the \DRbar and \MDRbar scheme.
All \DRbar soft-breaking mass parameters, the $\mu$ parameter of the
\MSSM super-potential, and the running \abbrev{CP}-odd Higgs mass are
set equal to $\MS$ here.  The running trilinear couplings, except
$A_t$, are chosen such that the sfermions do not mix.  The \DRbar stop
mixing parameter $X_t = A_t - \mu/\tan\beta$ is left as a free
parameter.
For this scenario we find that the difference between the \DRbar and
\MDRbar scheme is below $100\unit{MeV}$ for different values of the
stop mixing parameter.

Note that for all terms in the Higgs mass matrix except $\at$,
$\at\as$, and $\at\as^2$, it is perturbatively equivalent to use either
the \DRbar or the \MDRbar stop mass as defined above. Predominantly,
this concerns the electroweak contributions as well as the terms of
order $\at^2$. In this paper, we use the \DRbar stop mass for these
contributions.

\section{Implementation into \fs}\label{sec:fs}

\subsection{Determination of the \MSSM \DRbar parameters}
\label{sec:determination_of_MSSM_parameters}

\fs determines the running \DRbar gauge and Yukawa couplings as well
as the running vacuum expectation value of the \MSSM along the lines
of \citere{Pierce:1996zz} by setting the scale to the $Z$-boson pole
mass $M_Z$.  In this approach, the following Standard Model (\sm) input
parameters are used:
\begin{align}
  \begin{split}
    &\aem^{\SM(5)}(M_Z), \as^{\SM(5)}(M_Z),
    G_F, M_Z, \\
    &M_e, M_\mu, M_\tau, m_{u,d,s}(2\unit{GeV}), m_c^{\SM(4),\MSbar}(m_c), m_b^{\SM(5),\MSbar}(m_b),
    M_t \,,
  \end{split}
\end{align}
where $\aem^{\SM(5)}(M_Z)$ and $\as^{\SM(5)}(M_Z)$ denote the
electromagnetic and strong coupling constants in the \MSbar scheme in
the Standard Model with five active quark flavours, and $G_F$ is the
Fermi constant.  $M_e$, $M_\mu$, $M_\tau$, and $M_t$ denote the pole
masses of the electron, muon, tau lepton, and top quark, respectively.
The input masses of the up, down and strange quark are defined in the
\MSbar scheme at the scale $2\unit{GeV}$.  The charm and bottom quark
masses are defined in the \MSbar scheme at their scale in the Standard
Model with four and five active quark flavours, respectively.

The \MSSM \DRbar gauge couplings $g_1$, $g_2$ and $g_3$ are given in
terms of the \DRbar parameters $\aem^{\MSSM}(M_Z)$ and
$\as^{\MSSM}(M_Z)$ in the \MSSM as:
\begin{align}
  g_1(M_Z) &= \sqrt{\frac{5}{3}} \frac{\sqrt{4\pi\aem^{\MSSM}(M_Z)}}{\cos\theta_w(M_Z)} \,, \\
  g_2(M_Z) &= \frac{\sqrt{4\pi\aem^{\MSSM}(M_Z)}}{\sin\theta_w(M_Z)} \,, \\
  g_3(M_Z) &= \sqrt{4\pi\as^{\MSSM}(M_Z)} \,.
\end{align}
The couplings $\aem^{\MSSM}(M_Z)$ and $\as^{\MSSM}(M_Z)$ are calculated
from the corresponding input parameters as
\begin{align}
  \aem^{\MSSM}(M_Z) &=
  \frac{\aem^{\SM(5)}(M_Z)}{1 - \Delta\aem(M_Z)} \,,\\
  \as^{\MSSM}(M_Z) &=
  \frac{\as^{\SM(5)}(M_Z)}{1 - \Delta\as(M_Z)} \,,
\end{align}
where the threshold corrections $\Delta\alpha_i(M_Z)$ have the form
\begin{align}
  \Delta\aem(M_Z) &= \frac{\aem}{2\pi}
  \Bigg(\frac{1}{3}- \frac{16}{9} \log{\frac{m_{t}}{M_Z}} - \frac{4}{9}
  \sum_{i=1}^6 \log{\frac{m_{\tilde u_i}}{M_Z}} - \frac{1}{9}
  \sum_{i=1}^6 \log{\frac{m_{\tilde d_i}}{M_Z}} \nonumber \\
  &\phantom{={}} \qquad\quad -\frac{4}{3} \sum_{i=1}^2 \log{\frac{m_{\tilde \chi^+_i}}{M_Z}} -
  \frac{1}{3} \sum_{i=1}^6 \log \frac{m_{\tilde{e}_i}}{M_Z} -
  \frac{1}{3} \log{\frac{m_{H^+}}{M_Z}} \Bigg) \,, \\
  \Delta\as(M_Z) &= \frac{\as}{2\pi} \left[
    \frac{1}{2} - 2 \log{\frac{m_{\tilde g}}{M_Z}} - \frac{2}{3}
    \log{\frac{m_t}{M_Z}} - \frac{1}{6} \sum_{i=1}^6 \left(
      \log{\frac{m_{\tilde u_i}}{M_Z}} + \log{\frac{m_{\tilde
            d_i}}{M_Z}} \right) \right] \,.
  \label{eq:delta_alpha_s}
\end{align}
The \DRbar weak mixing angle in the \MSSM, $\theta_w$, is determined at
the scale $M_Z$ from the Fermi constant $G_F$ and the $Z$ pole mass via
the relation
\begin{align}
  \sin^2\theta_w \cos^2\theta_w =
  \frac{\pi\,\aem^{\MSSM}}{\sqrt{2} M_Z^2 G_F (1-\delta_r)} ,
\end{align}
where
\begin{align}
  \delta_r &= \hat\rho \frac{\re\Sigma_{W,T}(0)}{M_W^2} -
  \frac{\re\Sigma_{Z,T}(M_Z^2)}{M_Z^2} + \delta_{\abbrev{VB}} + \delta_r^{(2)} , \\
  \hat\rho &= \frac{1}{1-\Delta\hat\rho} ,\qquad\qquad \Delta\hat\rho
  = \re\Biggl[ \frac{\Sigma_{Z,T}(M_Z^2)}{\hat\rho\,M_Z^2} -
  \frac{\Sigma_{W,T}(M_W^2)}{M_W^2}\Biggr] + \Delta\hat\rho^{(2)} \,.
\end{align}
Here, $\Sigma_{V,T}(p^2)$ denotes the transverse part of the
\DRbar-renormalized \one-loop self energy of the vector boson $V$ in
the \MSSM.  The vertex and box
contributions $\delta_{\abbrev{VB}}$ as well as the
\two-loop contributions $\delta_r^{(2)}$ are taken from
\citere{Pierce:1996zz}.
The \DRbar vacuum expectation values of the up- and down-type Higgs
doublets are calculated by
\begin{align}
  v_u(M_Z) &= \frac{2 m_Z(M_Z) \sin\beta(M_Z)}{\sqrt{3/5 g_1^2(M_Z) + g_2^2(M_Z)}} \,, \\
  v_d(M_Z) &= \frac{2 m_Z(M_Z) \cos\beta(M_Z)}{\sqrt{3/5 g_1^2(M_Z) + g_2^2(M_Z)}} \,,
\end{align}
where $\tan\beta(M_Z)$ is an input parameter and $m_Z(M_Z)$ is the $Z$
boson \DRbar mass in the \MSSM, which is calculated from the $Z$ pole
mass at the \one-loop level as
\begin{align}
  m_Z^2(M_Z) = M_Z^2 + \re\Sigma_{Z,T}(M_Z^2) \,.
\end{align}

In order to calculate the Higgs pole mass in the \DRbar scheme at the
\three-loop level $\order{\atasas+\abasas}$, the \DRbar top and bottom Yukawa
couplings must be extracted from the input parameters $M_t$ and
$m_b^{\SM(5),\MSbar}(m_b)$ at the \two-loop level at $\order{\asas}$.  In
order to achieve that, we make use of the known \two-loop \SQCD
contributions to the top and bottom Yukawa couplings of
\citeres{Bednyakov:2002sf,Bednyakov:2005kt,Bednyakov:2007vm,Bauer:2008bj}, as
described in the following:
We calculate the \DRbar Yukawa couplings $y_t$ at the scale $M_Z$ from
the \DRbar top mass $m_t$ and the \DRbar up-type \abbrev{VEV} $v_u$ as
\begin{align}
  y_t(M_Z) = \sqrt{2} \frac{m_t(M_Z)}{v_u(M_Z)} \,.
\end{align}
In our approach, we relate the \DRbar top mass to the top pole mass
$M_t$ at the scale $M_Z$ as
\begin{align}
\begin{split}
  m_t(M_Z) &= M_t + \re\Sigma_{t}^S(M_t^2,M_Z) \\
  &\phantom{={}} + M_t \Big[ \re\Sigma_{t}^L(M_t^2,M_Z) +
    \re\Sigma_{t}^R(M_t^2,M_Z) \\
  &\phantom{={} + M_t \Big[}
    + \Delta m_t^{(1),\abbrev{SQCD}}(M_Z) + \Delta m_t^{(2),\abbrev{SQCD}}(M_Z) \Big]
  \,,
\end{split}
\label{eq:mt_FS}
\end{align}
where $\Sigma_{t}^{S,L,R}(p^2,Q)$ denote the scalar (superscript $S$),
and the left- and right-handed parts $(L,R)$ of the \DRbar
renormalized \one-loop top self-energy without the gluon, stop, and
gluino contributions, and $\Delta m_t^{(1),\abbrev{SQCD}}$ and $\Delta
m_t^{(2),\abbrev{SQCD}}$ are the full
\one- and \two-loop \SQCD corrections taken from
\citeres{Bednyakov:2002sf,Bednyakov:2005kt},
{\allowdisplaybreaks\begin{align}
  \Delta m_t^{(1),\text{SQCD}} &=
  - \frac{\as}{4 \pi} C_F \Bigg[
   \Bigg(\frac{m_g m_{\tilde{t}_1}^2 s_{2\theta_t}}{m_t
   (m_{\tilde{t}_1}^2-m_g^2)}-\frac{m_g m_{\tilde{t}_2}^2
   s_{2\theta_t}}{m_t
   (m_{\tilde{t}_2}^2-m_g^2)}+\frac{m_{\tilde{t}_1}^4}{2
   (m_{\tilde{t}_1}^2-m_g^2)^2} \nonumber \\
   &\phantom{={}}\qquad -\frac{m_{\tilde{t}_1}^2}{m_{\tilde{t}_1}^2-m_g^2}+\frac{m_{\tilde{t}_2}^4}{2
   (m_{\tilde{t}_2}^2-m_g^2)^2}-\frac{m_{\tilde{t}_2}^2}{m_{\tilde{t}_2}^2-m_g^2}+1\Bigg)\log\frac{m_g^2}{Q^2} \nonumber \\
   &\phantom{={}}\qquad + \Bigg(-\frac{m_g m_{\tilde{t}_1}^2 s_{2\theta_t}}{m_t
   (m_{\tilde{t}_1}^2-m_g^2)}-\frac{m_{\tilde{t}_1}^4}{2
   (m_{\tilde{t}_1}^2-m_g^2)^2}+\frac{m_{\tilde{t}_1}^2}{m_{\tilde{t}_1}^2-m_g^2}\Bigg) \log\frac{m_{\tilde{t}_1}^2}{Q^2} \nonumber \\
   &\phantom{={}}\qquad + \Bigg(\frac{m_g m_{\tilde{t}_2}^2 s_{2\theta_t}}{m_t
   (m_{\tilde{t}_2}^2-m_g^2)}-\frac{m_{\tilde{t}_2}^4}{2
   (m_{\tilde{t}_2}^2-m_g^2)^2}+\frac{m_{\tilde{t}_2}^2}{m_{\tilde{t}_2}^2-m_g^2}\Bigg) \log\frac{m_{\tilde{t}_2}^2}{Q^2} \nonumber \\
   &\phantom{={}}\qquad +\frac{m_{\tilde{t}_1}^2}{2
   (m_{\tilde{t}_1}^2-m_g^2)}+\frac{m_{\tilde{t}_2}^2}{2
   (m_{\tilde{t}_2}^2-m_g^2)}-3 \log\frac{m_t^2}{Q^2}+\frac{7}{2}
  \Bigg] \,,
  \label{eq:dmt_SQCD_1L} \\
  \Delta m_t^{(2),\text{SQCD}} &= \left(\Delta m_t^{(1),\text{SQCD}}\right)^2
  - \Delta m_t^{(2),\text{dec}} \,.
  \label{eq:dmt_SQCD_2L}
\end{align}}%
In \eqn{eq:dmt_SQCD_1L}, it is $C_F = 4/3$ and $s_{2\theta_t} = \sin
2\theta_t$, with $\theta_t$ the stop mixing angle.  The \two-loop term
$\Delta m_t^{(2),\text{dec}}$ is given in \citere{Bednyakov:2002sf} for
general stop, sbottom, and gluino masses.

The \MSSM \DRbar bottom-quark Yukawa coupling $y_b$ is calculated from
the \DRbar bottom-quark mass $m_b$ and the down-type \abbrev{VEV} at the scale
$M_Z$ as
\begin{align}
  y_b(M_Z) = \sqrt{2} \frac{m_b(M_Z)}{v_d(M_Z)} \,.
\end{align}
We obtain $m_b(M_Z)$ from the input \MSbar mass $m_b^{\SM(5),\MSbar}(m_b)$ in the
Standard Model with five active quark flavours by first evolving
$m_b^{\SM(5),\MSbar}(m_b)$ to the scale $M_Z$, using the \one-loop \QED and
\three-loop \QCD\ renormalization group equations (\abbrev{RGE}s).
Afterwards, $m_b^{\SM(5),\MSbar}(M_Z)$
is converted to the \DRbar mass $m_b^{\SM(5),\DRbar}(M_Z)$ by the
relation
\begin{align}
  m_b^{\SM(5),\DRbar}(M_Z) = m_b^{\SM(5),\MSbar}(M_Z) \left(1 -
    \frac{\as}{3 \pi} + \frac{3 g_2^2}{128 \pi^2} +
    \frac{13 g_Y^2}{1152 \pi^2}\right) \,.
\end{align}
Finally, the \MSSM \DRbar bottom mass $m_b(M_Z)$ is obtained from
$m_b^{\SM(5),\DRbar}(M_Z)$ via
\begin{align}
  m_b(M_Z) &= \frac{m_b^{\SM(5),\DRbar}(M_Z)}{1 + \Delta m_b^{(1)} + \Delta m_b^{(2)}} \,, \\
  \Delta m_b^{(1)} &= -\re\Sigma_{b}^S((m_b^{\SM(5),\MSbar})^2,M_Z)/m_b \notag\\
  &\phantom{={}}- \re\Sigma_{b}^L((m_b^{\SM(5),\MSbar})^2,M_Z) -
  \re\Sigma_{b}^R((m_b^{\SM(5),\MSbar})^2,M_Z) \,, \\
  \label{eq:bottom-conversion}
  \Delta m_b^{(2)} &= \Delta m_b^{(2),\text{dec}}
  - \frac{\as}{3\pi} \Delta m_b^{(1)} \,,
\end{align}
where $\Sigma_{b}^{S,L,R}(p^2,Q)$ are the scalar, left- and right-handed
parts of the \DRbar renormalized \one-loop bottom quark self-energy in
the \MSSM, in which all Standard Model particles, except the bottom
quark, the top quark and the $W$, $Z$, and Higgs bosons, are omitted.
In \eqn{eq:bottom-conversion} $\Delta m_b^{(2),\text{dec}}$ denotes
the \two-loop decoupling relation of order $\order{\asas}$ between
the \MSbar bottom mass $m_b^{\SM(5),\MSbar}$ and the \DRbar bottom mass
in the \MSSM calculated in
\citeres{Bednyakov:2007vm,Bauer:2008bj}.

Note that the matching of the \SM to the \MSSM leads to large
logarithmic contributions in the \MSSM \DRbar parameters in the case of
a heavy \SUSY particle spectrum.  These contributions can be resummed in
a so-called \EFT
approach\,\cite{Bagnaschi:2014rsa,Vega:2015fna,Athron:2016fuq,Bahl:2016brp,
Bahl:2017aev}.

\subsection{Calculation of the \abbrev{CP}-even Higgs pole masses}

\fs calculates the two \abbrev{CP}-even Higgs pole masses $M_h$ and $M_H$ by
diagonalizing the loop-corrected mass matrix\footnote{We do not
distinguish between \DRbar and \MDRbar parameters here, and drop the hat
over $\hat{\mathsf{M}}$ introduced in \eqn{eq:Mmdr} for simplicity.}
\begin{align}
  \mathsf{M} = \mathsf{M}^\tree + \mathsf{M}^{1L}(p^2)
  + \mathsf{M}^{2L} + \mathsf{M}^{3L}
  \label{eq:Mh_diagonalization}
\end{align}
at the momenta $p^2 = M_h^2$ and $p^2 = M_H^2$, respectively
($\mathsf{M}^{2L}$ and $\mathsf{M}^{3L}$ are evaluated at $p^2=0$).  The
\one-loop correction $\mathsf{M}^{1L}(p^2)$ contains the full \one-loop
\MSSM Higgs self energy and tadpole contributions, including electroweak
corrections and the momentum dependence.  The \two-loop correction
$\mathsf{M}^{2L}$ contains the known corrections of order $\order{\as
(\at + \ab) + (\at+\ab)^2 + \atau^2}$
\cite{Degrassi:2001yf,Brignole:2001jy,Dedes:2002dy,Brignole:2002bz,Dedes:2003km}.
The \three-loop correction $\mathsf{M}^{3L}$ incorporates the terms of
order $\order{\atasas + \abasas}$ from the \himalaya package, as described
in \secref{sec:higgsMass3L}.  In
Eq.~\eqref{eq:Mh_diagonalization} all contributions are defined in the
\DRbar scheme by default.\footnote{\fsh provides a flag to calculate
  the corrections of order
  $\mathcal{O}(\at(1 + \as + \as^2) + \ab(1 + \as + \as^2))$ in the \MDRbar
  scheme, as described in Section~\ref{sec:MDR_scheme}.  See
  Appendix~\ref{app:FS_options} for more details.}
The renormalization scale is chosen to be
$Q = \sqrt{\mstop{1}\mstop{2}}$ and the \DRbar parameters which enter
Eq.~\eqref{eq:Mh_diagonalization} are evolved to that scale by using
the \three-loop \abbrev{RGE}s of the \MSSM\,\cite{Jack:2003sx,Jack:2004ch}.
Since the two \abbrev{CP}-even Higgs pole masses are the output of the
diagonalization of $\mathsf{M}$ but at the same time must be inserted
into $\mathsf{M}^{1L}(p^2)$, an iteration over the momentum is performed
for each mass eigenvalue until a fixed point for the Higgs masses is
reached with sufficient precision.

\section{Results}\label{sec:results}

\subsection{Size of \three-loop contributions from different
  sources}\label{sec:3loop}

\begin{figure}[tbh]
  \centering
  \includegraphics[width=0.49\textwidth]{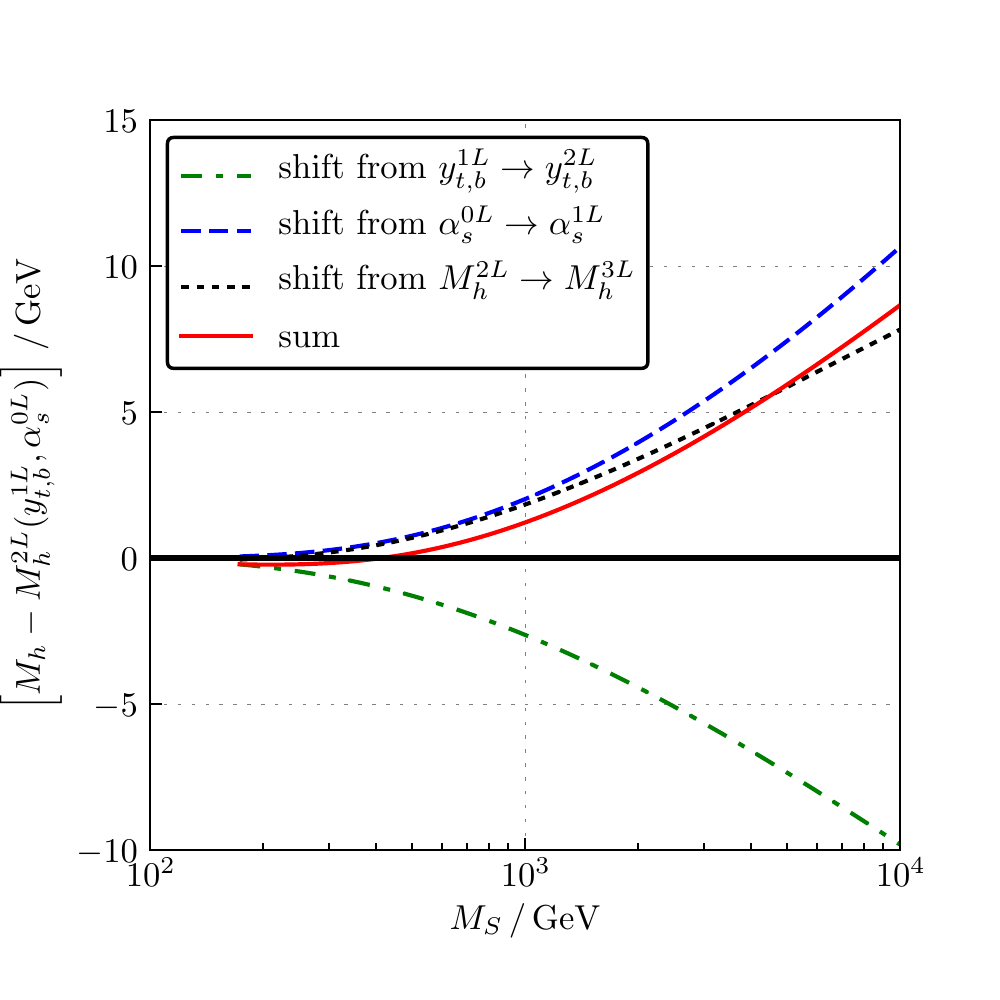}\hfill
  \includegraphics[width=0.49\textwidth]{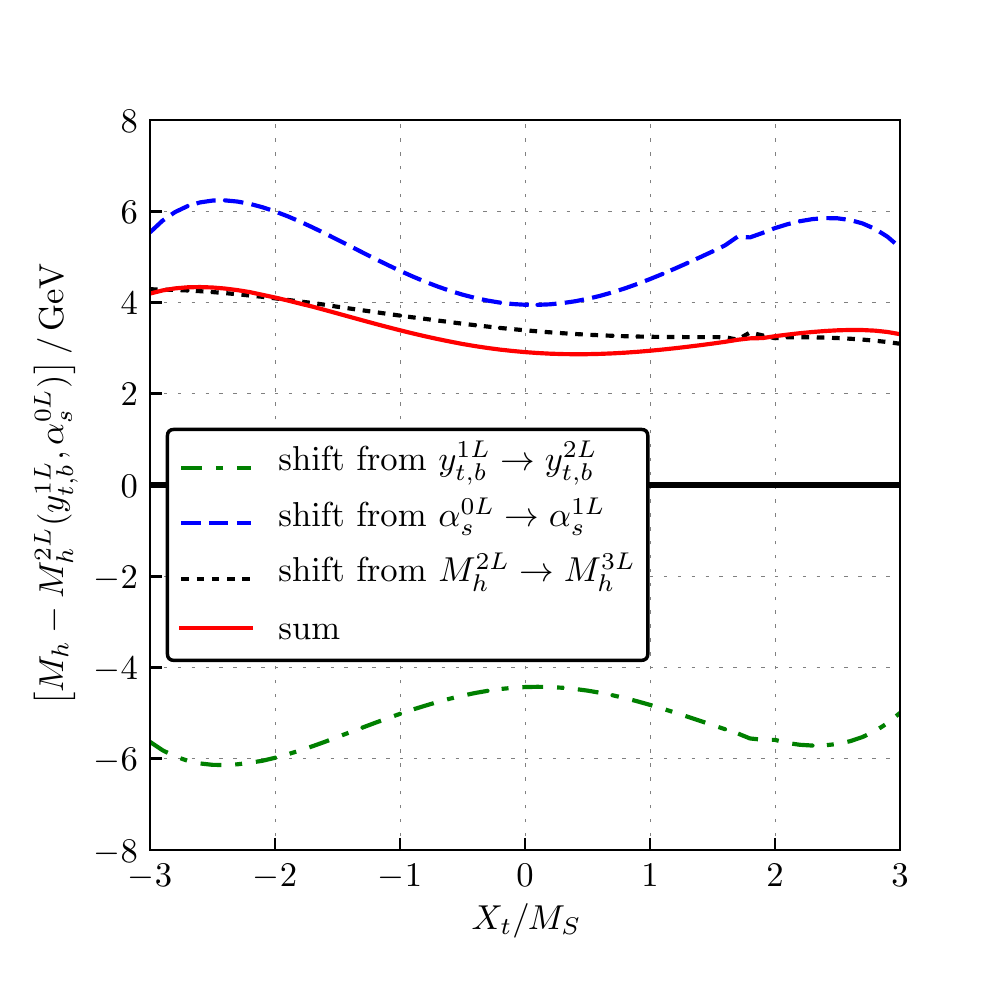}
  \caption{Influence of different \three-loop contributions to the Higgs
    pole mass.  In the left panel we show the shift in the Higgs pole
    mass with respect to $M_h^{2L}(y_{t,b}^{1L},\as^{0L})$ for
    $\tan\beta = 5$ and $X_t = 0$ as a function of the \SUSY scale.  In
    the right panel we fix $\tan\beta = 5$ and $\MS = 2\unit{TeV}$ and
    vary the relative stop mixing parameter $X_t/\MS$.}
  \label{fig:Mh_MS_Xt_contributions}
\end{figure}

In the \DRbar calculation within \fsh, there are three sources of
contributions which affect the Higgs pole mass at order $\order{\atasas
  + \abasas}$: The \one-loop threshold correction $\order{\as}$ to the
strong coupling constant, the \two-loop threshold correction
$\order{\asas}$ to the top and bottom Yukawa couplings, and the genuine
\three-loop contribution to the Higgs mass matrix.  In
\figref{fig:Mh_MS_Xt_contributions}, the impact of these three sources
on the Higgs pole mass is shown relative to the \two-loop calculation
without these three corrections.  The left panel shows the impact as a
function of the \SUSY scale $\MS$, and the right panel as a function of
the relative stop mixing parameter $X_t/\MS$ for the scenario defined
in \secref{sec:MDR_scheme}.

First, we observe that the inclusion of the \one-loop threshold
correction to $\as$, \eqn{eq:delta_alpha_s}, (blue dashed line)
leads to a significant positive shift of the Higgs pole mass of around
$+2.5\unit{GeV}$ for $\MS\approx 1\unit{TeV}$.  For larger \SUSY scales
the shift increases logarithmically as is to be expected from the
logarithmic terms on the r.h.s.\ of \eqn{eq:delta_alpha_s}.
The inclusion of the full \two-loop \SQCD corrections to $y_t$ (green
dash-dotted line) leads to a shift of similar magnitude, but in the
opposite direction (the effect due to $y_b$ is negligible).  Thus, there
is a significant cancellation between the \three-loop contributions from
the \one-loop threshold correction to $\as$ and the \two-loop \SQCD
corrections to $y_t$.
The genuine \three-loop contribution to the Higgs pole mass (black
dotted line) is again positive and around $+2\unit{GeV}$ for $\MS\approx
1\unit{GeV}$.  This is consistent with the findings of
\citere{Kant:2010tf}, of course.
As a result, the sum of these three \three-loop effects (red solid line)
leads to a net positive shift of the Higgs mass relative to
the \two-loop result without all these corrections.

The size of the individual \three-loop contributions depends on the stop
mixing parameter $X_t/\MS$, as can be seen from the r.h.s.\ of
\figref{fig:Mh_MS_Xt_contributions}: between minimal ($X_t/\MS = 0$) and
maximal stop mixing ($X_t/\MS\approx\sqrt{6}$) the size of the
individual \three-loop contributions changes by $1$--$2\unit{GeV}$.  For
maximal (minimal) mixing, their impact is maximal (minimal). The
direction of the shift is independent of $X_t/\MS$.

Note that the nominal \two-loop result of the original \fs (i.e.,
without \himalaya) includes by default the \one-loop threshold
correction to $\as$ and the \sm\ \QCD \two-loop contributions to the top
Yukawa coupling \cite{Athron:2014yba,Athron:2016fuq}.  This
means that the \two-loop Higgs mass as evaluated by the original \fs
already incorporates partial \three-loop contributions.  As a result,
the \two-loop result of the original \fs does not correspond to the
zero-line in \figref{fig:Mh_MS_Xt_contributions}, but is rather close to
the blue dashed line.  This implies that, compared to the \two-loop
result of the original \fs, the effect of the remaining $\at\as^2$
contributions in the Higgs mass prediction is \emph{negative}.

\subsection{Scale dependence of the \three-loop Higgs pole mass}

\begin{figure}[tbh]
  \centering
  \includegraphics[width=0.49\textwidth]{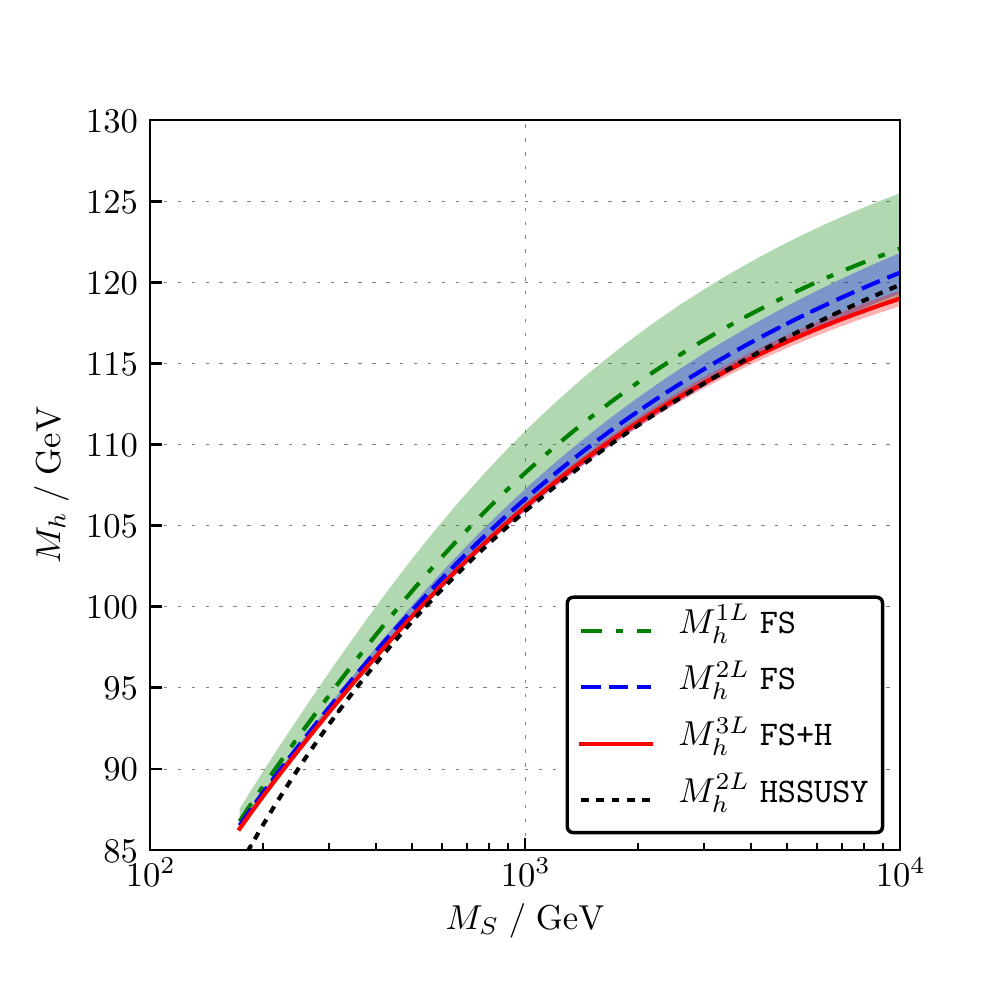}\hfill
  \includegraphics[width=0.49\textwidth]{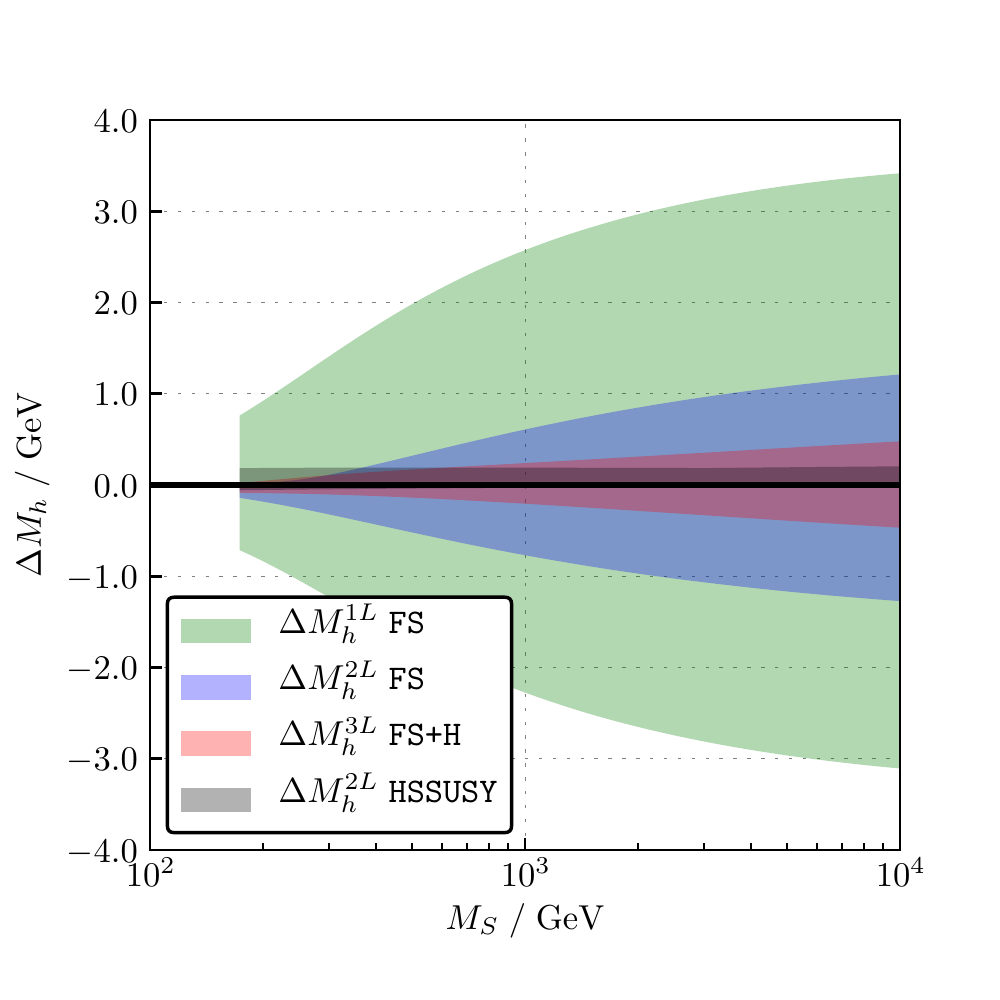}
  \caption{Variation of the Higgs pole mass when the renormalization
    scale is varied by a factor two at which the Higgs pole mass is
    calculated, for $\tan\beta = 5$ and $X_t = 0$.}
  \label{fig:Mh_MS_Xt_uncertainty}
\end{figure}

To estimate the size of the missing higher-order corrections,
\figref{fig:Mh_MS_Xt_uncertainty} shows the renormalization scale
dependence of the \one-, \two- and \three-loop Higgs pole mass for the
scenario defined in \secref{sec:MDR_scheme} with $\tan\beta = 5$ and
$X_t = 0$.
The \one- and \two-loop calculations correspond to the original \fs.  In
the \one-loop calculation the threshold corrections to $\as$ and $y_t$
are set to zero, and in the \two-loop calculation the \one-loop
threshold corrections to $\as$ and the \two-loop \QCD corrections to
$y_t$ are taken into account.  The \three-loop result of \fsh includes
all \three-loop contributions at $(\at+\ab)\as^2$ discussed above,
i.e.\ the \one-loop threshold correction to $\as$, the full \two-loop
\SQCD corrections to $y_{t,b}$, and the genuine \three-loop correction
to the Higgs pole mass from \himalaya.  In addition, the Higgs mass
predicted at the \two-loop level in the pure \EFT calculation of \HSSUSY
is shown as the black dotted line, see
\secref{sec:comparison_with_others}.
The bands show the corresponding variation of the Higgs pole mass when
the renormalization scale is varied using the \three-loop
renormalization group equations
\cite{Jack:2003sx,Jack:2004ch,Mihaila:2012fm,Bednyakov:2012rb,Bednyakov:2012en,Chetyrkin:2012rz,Bednyakov:2013eba}
for all parameters except for the vacuum expectation values, where the
$\beta$-functions are known only up to the \two-loop level
\cite{Sperling:2013eva,Sperling:2013xqa}.  In \fs and \fsh, the
renormalizaion scale is varied in the full \MSSM within the interval
$[\MS/2,2\MS]$, while in \HSSUSY it is varied in the Standard Model
within the interval $[M_t/2,2 M_t]$, keeping the matching scale fixed at
$\MS$. The plot shows that the successive
inclusion of higher-order corrections reduces the scale dependence, as
expected.  In particular, the \three-loop corrections to the Higgs mass
reduce the scale dependence by around a factor two, compared to the
\two-loop calculation.  The scale dependence of \HSSUSY is almost
independent of \MS, because scale variation is done within the
\SM\ after integrating out all SUSY particles at \MS.
Note that the variation of the renormalization scale only serves as an
indicator of the theoretical uncertainty due to missing higher order
effects.

\subsection{Comparison with lower order and \EFT results}
\label{sec:comparison_with_others}

\begin{figure}[tbh]
  \centering
  \includegraphics[width=0.49\textwidth]{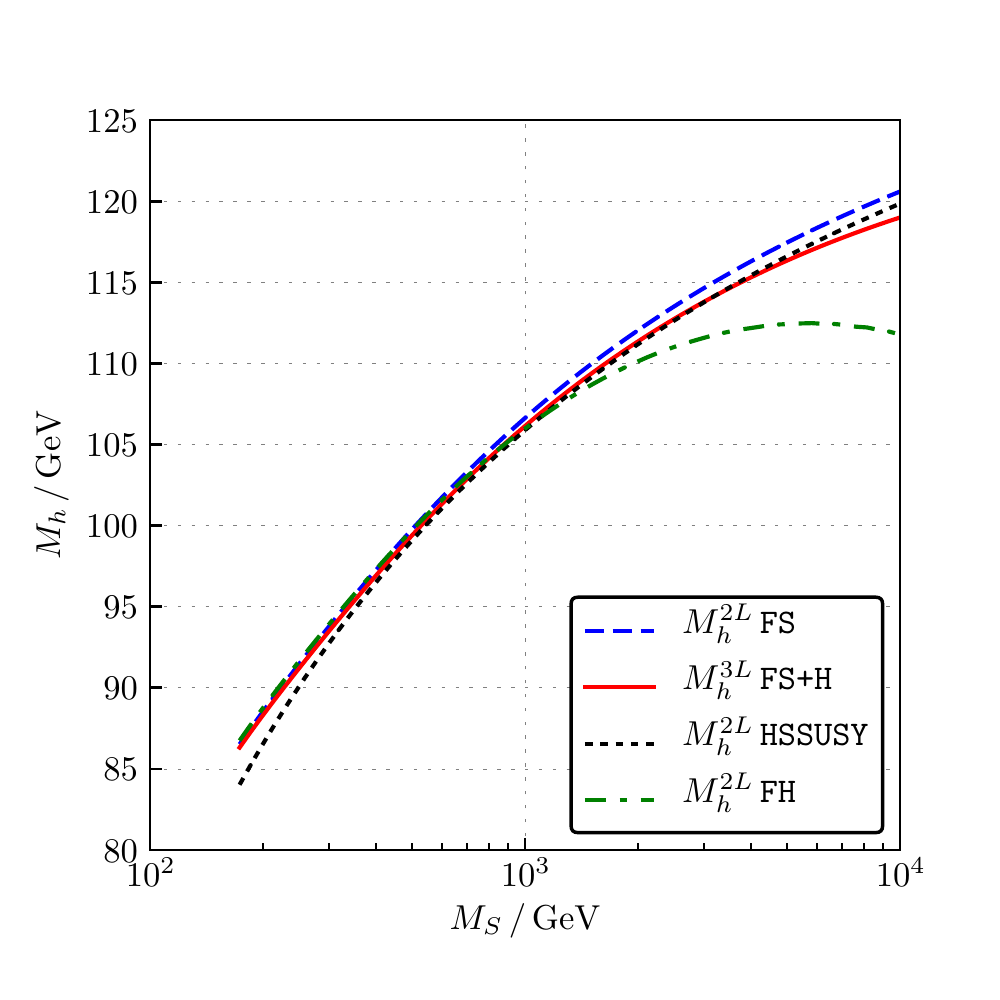}\hfill
  \includegraphics[width=0.49\textwidth]{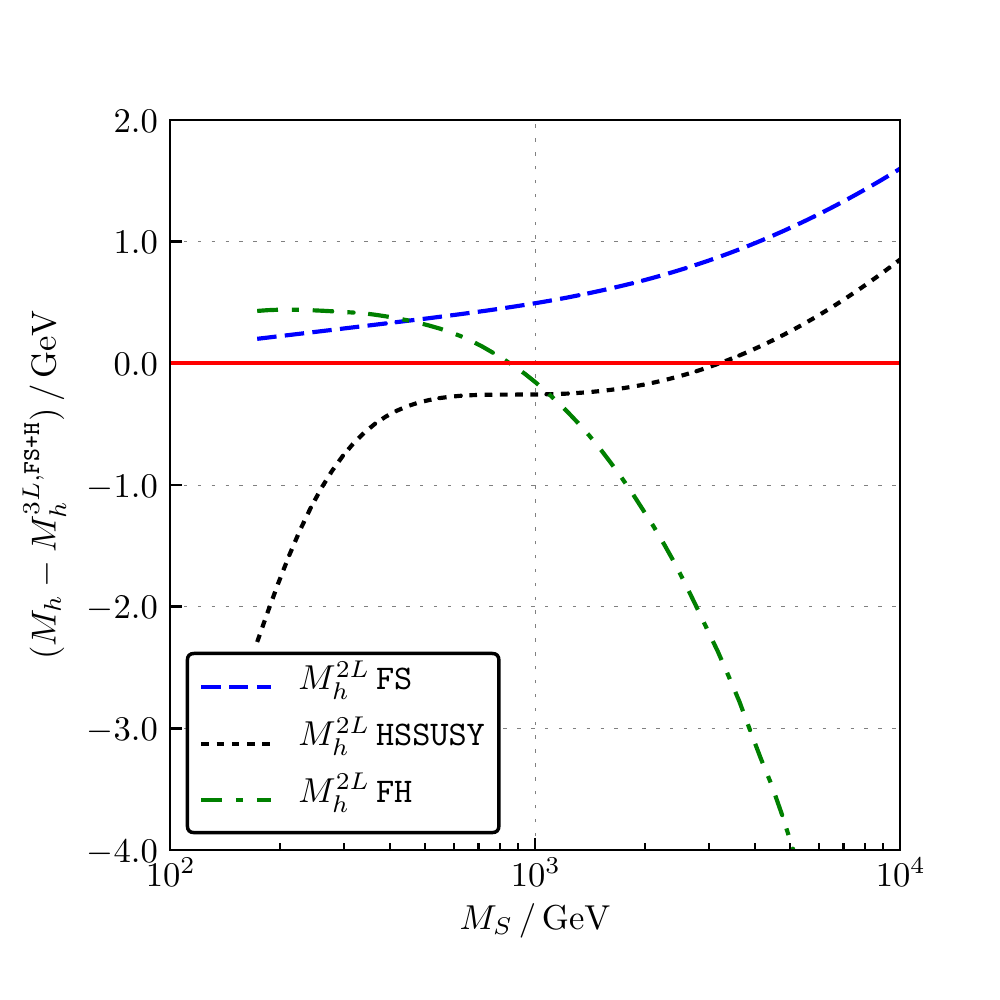}
  \caption{Comparison of Higgs mass predictions between \two- and
    \three-loop fixed-order programs and a \two-loop \EFT calculation as
    a function of the \SUSY scale for $\tan\beta = 5$ and $X_t = 0$.  In
    the left panel the absolute Higgs pole mass and in the right panel
    the difference w.r.t.\ the \three-loop calculation is
    shown (\texttt{FS}=\fs, \texttt{FS+H}=\fsh,
    \texttt{FH}=\FeynHiggs).}
  \label{fig:Mh_MS}
\end{figure}

\begin{figure}[tbh]
  \centering
  \includegraphics[width=0.49\textwidth]{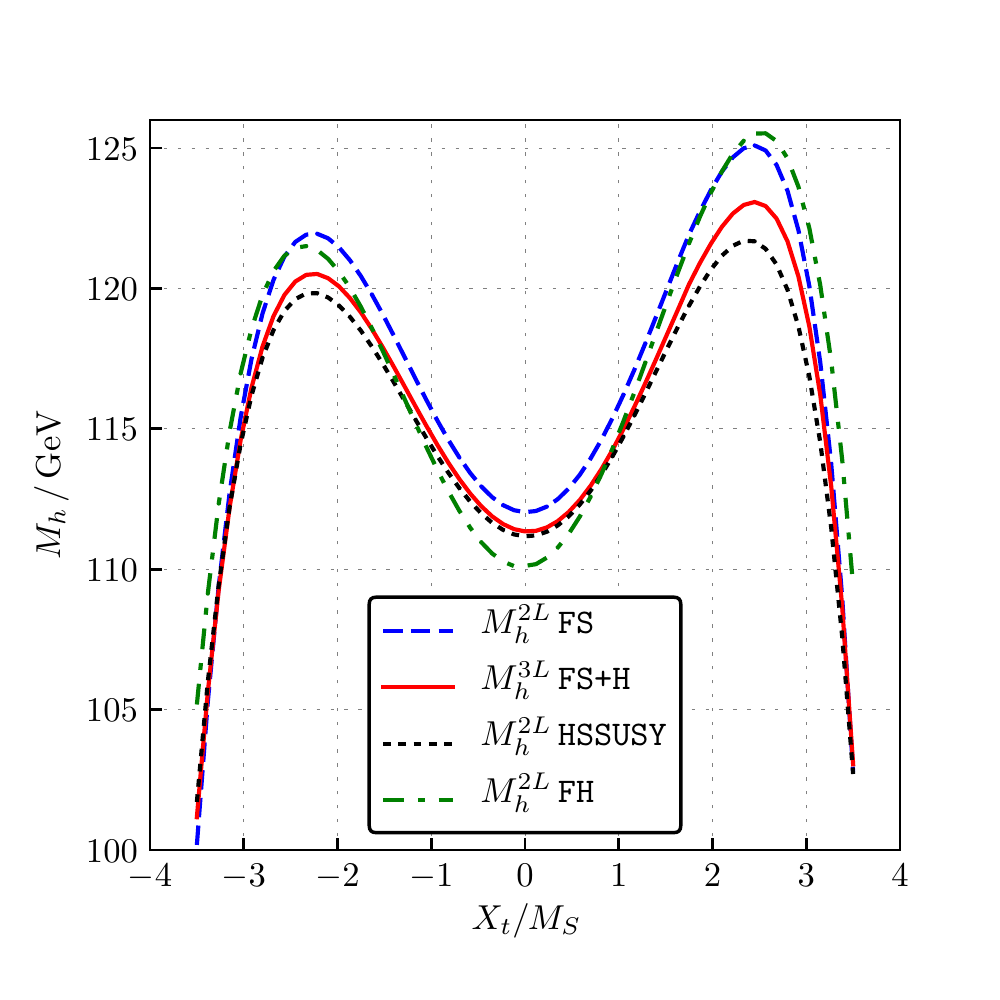}\hfill
  \includegraphics[width=0.49\textwidth]{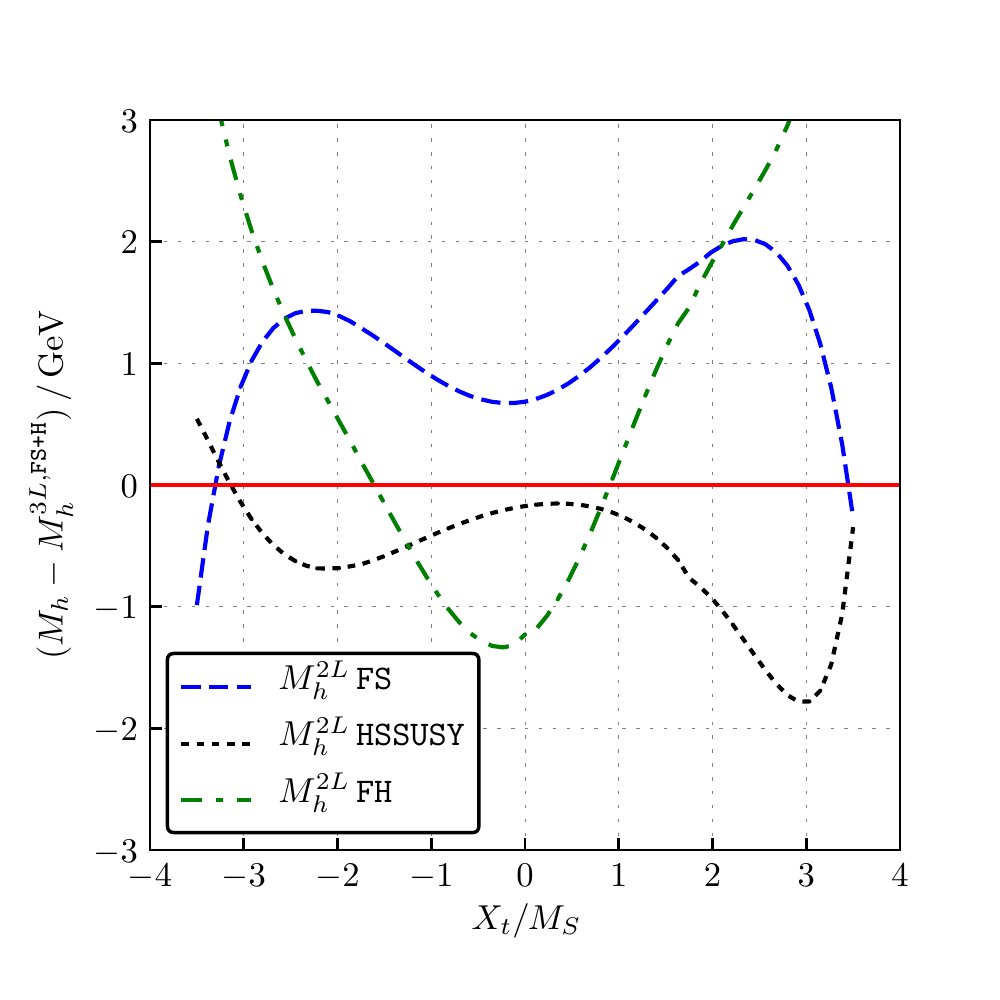}
  \caption{Comparison of Higgs mass predictions between \two- and \three-loop
    fixed-order programs and a \two-loop \EFT calculation as a function of
    the relative stop mixing parameter $X_t/\MS$ for $\tan\beta = 5$
    and $\MS = 2\unit{TeV}$.  In the left panel the absolute Higgs
    pole mass and in the right panel the difference w.r.t.\ the
    \three-loop calculation is shown.}
  \label{fig:Mh_Xt}
\end{figure}

In \figsref{fig:Mh_MS}--\ref{fig:Mh_Xt}, we compare the \three-loop
calculation of \fsh (red) with other \MSSM spectrum generators.  As
input we use $M_t = 173.34\unit{GeV}$, $\aem^{\SM(5)}(M_Z) =
1/127.95$, $\as^{\SM(5)}(M_Z) = 0.1184$ and $G_F = 1.1663787\cdot
10^{-5} \unit{GeV}^{-2}$.
All \DRbar soft-breaking mass parameters as well as the $\mu$ parameter
of the super-potential in the \MSSM, and the running \abbrev{CP}-odd
Higgs mass are set equal to $\MS$.  The running trilinear couplings,
except for $A_t$, are chosen such that there is no sfermion mixing.  The
stop mixing parameter $X_t = A_t - \mu/\tan\beta$ is defined in the
\DRbar scheme and left as a free parameter.  The lightest
\abbrev{CP}-even Higgs pole mass is calculated at the scale $Q =
\sqrt{\mstop{1}\mstop{2}}$.

\paragraph{FlexibleSUSY 1.7.4.}
The blue dashed line shows the original \two-loop calculation with \fs
1.7.4\,\cite{Athron:2014yba}. Note that, by construction of \fs, this
result coincides exactly with the one of \softsusy 3.5.1.  As described
above, it includes the \one-loop threshold corrections to $\as$ and the
\two-loop \QCD contributions to $y_t$, and it uses the \three-loop
\abbrev{RGEs} of the \MSSM\,\cite{Jack:2003sx,Jack:2004ch}.  \fs 1.7.4
(and \softsusy) use the explicit \two-loop Higgs pole mass
contribution of order $\order{\as (\at + \ab) + (\at+\ab)^2 + \atau^2}$
of
\citeres{Degrassi:2001yf,Brignole:2001jy,Dedes:2002dy,Brignole:2002bz,Dedes:2003km}.

\paragraph{HSSUSY 1.7.4.}
The black dotted line has been obtained using the pure \two-loop
effective field theory (\EFT) calculation of
\HSSUSY\,\cite{flexiblesusy}.  \HSSUSY is a spectrum generator from the \fs
suite, which implements the \two-loop threshold correction for the
quartic Higgs coupling of the Standard Model at $\order{\at (\at +
  \as)}$ when integrating out the \SUSY particles at a common \SUSY
scale \cite{Bagnaschi:2014rsa,Vega:2015fna}.  Renormalization group
running is performed down to the top mass scale using the \three-loop
\abbrev{RGEs} of the Standard Model
\cite{Mihaila:2012fm,Bednyakov:2012rb,Bednyakov:2012en,Chetyrkin:2012rz,Bednyakov:2013eba}
and finally the Higgs mass is calculated at the \two-loop level in the
Standard Model at order $\order{\at (\at + \as)}$.  In terms of the
implemented corrections, \HSSUSY is equivalent to
\susyhd\,\cite{Vega:2015fna}, and resums large logarithms up to
\abbrev{NNLL} level while neglecting terms of order $v^2/\MS^2$.
The $\order{v^2/\MS^2}$ corrections calculated in
\citere{Bagnaschi:2017xid} have not been taken into account here.

\paragraph{FeynHiggs 2.13.0-beta.}
The green dash-dotted line shows the Higgs mass prediction using
\FeynHiggs 2.13.0-beta without large log resummation
\cite{Heinemeyer:1998yj,Heinemeyer:1998np,Degrassi:2002fi,Frank:2006yh,Hahn:2013ria,Bahl:2016brp}.%
\footnote{We use the \abbrev{SLHA} input interface of \FeynHiggs, which
  performs a conversion of the \DRbar input parameters to the on-shell
  scheme.  Resummation is disabled, as it would lead to an
  inconsistent result in combination with the \DRbar to on-shell
  conversion of \FeynHiggs \cite{Bahl:2017aev}.  We call \FeynHiggs
  with the flags 4002020110.}
\FeynHiggs 2.13.0-beta includes the \two-loop contributions of
order $\order{\at\as + \ab\as + \at^2 + \at\ab}$.
\vspace{\baselineskip}

Consider first \figref{fig:Mh_MS}. The left panel shows the Higgs mass
prediction as a function of $\MS$ according to three codes discussed
above, together with the \fsh result (solid red). The stop mixing
parameter $X_t$ is set to zero. The right panel shows the difference of
these curves to the latter. Note that the resummed result of \HSSUSY
neglects terms of order $v^2/\MS^2$, and thus forfeits reliability
towards lower values of $\MS$. The deviation from the fixed order curves
below $\MS\approx 400$\,GeV clearly underlines this. In contrast, the
fixed order results start to suffer from large logarithmic contributions
toward large $\MS$, which on the other hand are properly resummed in the
\HSSUSY approach.  From \figref{fig:Mh_MS}, we conclude that the
fixed-order \DRbar result loses its applicability once $\MS$ is larger
than a few TeV, while the deviation between the non-resummed on-shell result of
\FeynHiggs and \HSSUSY increases more rapidly above $\MS\approx 1$\,TeV.
Note that the good agreement of \fs with \HSSUSY above the few-TeV
region is accidental, as shown in \citere{Athron:2016fuq}.

The effect of the \three-loop $\at\as^2$ terms on the fixed-order result
is negative, as discussed in \secref{sec:3loop}, and amounts to a
few hundred MeV in the region where the fixed-order approach is
appropriate. They significantly improve the agreement between the
fixed-order and the resummed prediction for $M_h$ in the intermediate
region of $\MS$, where both approaches are expected to be
reliable. Between $\MS$ of about 500\,GeV and 5\,TeV, our \three-loop
curve from \fsh deviates from the \HSSUSY result by less than
300\,MeV. This corroborates the compatibility of the two approaches in
the intermediate region. Considering the current estimate of the
theoretical uncertainty in the Higgs mass prediction
\cite{Degrassi:2002fi,Allanach:2004rh,Bagnaschi:2014rsa,Vega:2015fna,Athron:2016fuq}, our
observation even legitimates a naive switching between the fixed-order
and the resummed approach at $\MS\approx 1$\,TeV, instead of a more
sophisticated matching procedure along the lines of
\citere{Bahl:2016brp,Bahl:2017aev}. Nevertheless, the latter is clearly
desirable through order $\at\as^2$, in particular in the light of the
observations for non-zero stop mixing to be discussed below, but has to
be deferred to future work at this point.

\begin{figure}[tbh]
  \centering \includegraphics[width=0.49\textwidth]{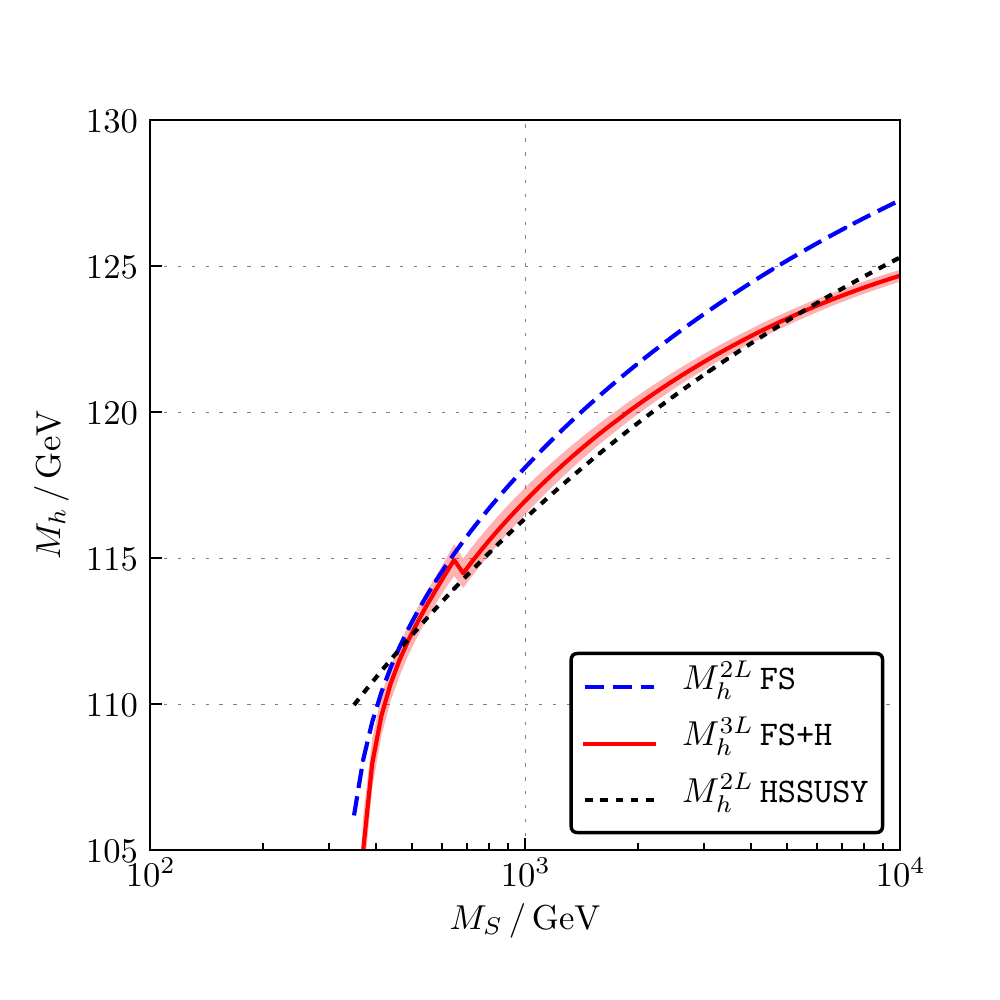}
  \caption{Comparison of the lightest Higgs pole mass calculated at
    the \two- and \three-loop level with \fs, \fsh and \HSSUSY as a
    function of the \SUSY scale $\MS$ for $\tan\beta = 5$ and
    $X_t = -\sqrt{6}\MS$.  The red band shows the size of the
    hierarchy selection criterion $\delta_i$.  In the fixed-order
    calculations of \fs and \fsh the Higgs mass becomes tachyonic
    for $\MS \lesssim 350\unit{GeV}$.}
  \label{fig:Mh_MS_TB-5_Xt--sqrt6}
\end{figure}

\figref{fig:Mh_Xt} shows the \three-loop effects as a function of $X_t$,
where the value of $\MS=2$\,TeV is chosen to be inside the intermediate
region. The figure shows that, for $|X_t|\lesssim 3\MS$, the qualitative
features of the discussion above are largely independent of the mixing
parameter, whereupon the quantitative differences between the
fixed-order and the resummed results are typically larger for non-zero
stop mixing. \figref{fig:Mh_MS_TB-5_Xt--sqrt6} underlines this by
setting $X_t=-\sqrt{6}\MS$ and varying $\MS$.  The kink
in the \three-loop curve originates from a change of the optimal
hierarchy chosen by \himalaya.  The red band shows the uncertainty
$\delta_i$ as defined in Eq.~\eqref{eq:deltai-definition}, which is used
to select the best fitting hierarchy.  We find that $\delta_i$ is
comparable to the size of the kink, which indicates a reliable treatment
of the hierarchy selection criterion.

\subsection{Comparison with other \three-loop results}

The \three-loop $\order{\at\as^2}$ corrections to the light \mssm Higgs
mass discussed in this paper were originally implemented in the
\mathematica code \htm.
We checked that the implementation of the $\at$ and $\at\as$ terms
in \himalaya leads to the same numerical results as in \htm, if the same
set of \DRbar parameters is used as input. Since the $\at\as^2$ terms
of \himalaya are derived from their implementation in \htm, it is not
surprising that they also result in the same numerical value if the same
set of input parameters is given \textit{and} the same mass hierarchy is
selected. But since \himalaya has a slightly more sophisticated way of
choosing this hierarchy (see \secref{sec:hierarchy}), its
numerical $\at\as^2$ contribution does occasionally differ slightly from the one
of \htm.

\begin{figure}[tbh]
  \centering
  \includegraphics[width=0.49\textwidth]{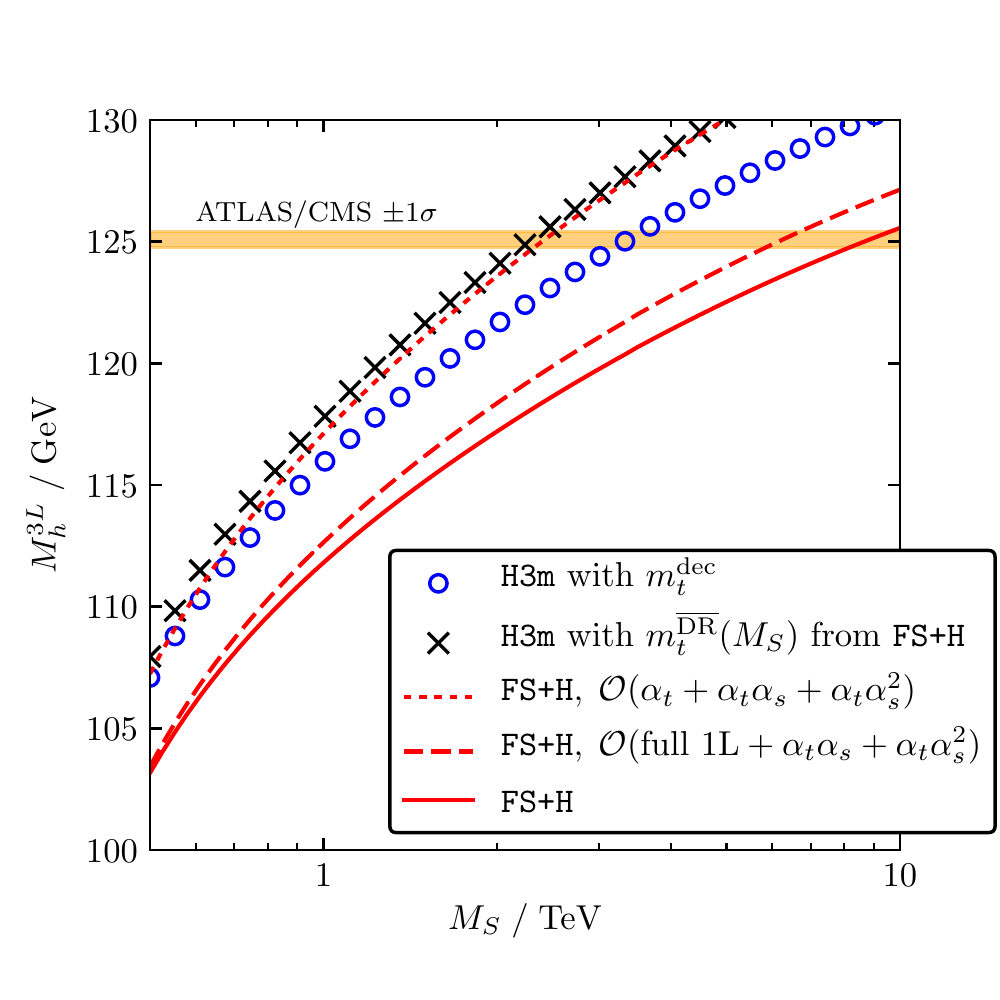}\hfill
  \includegraphics[width=0.49\textwidth]{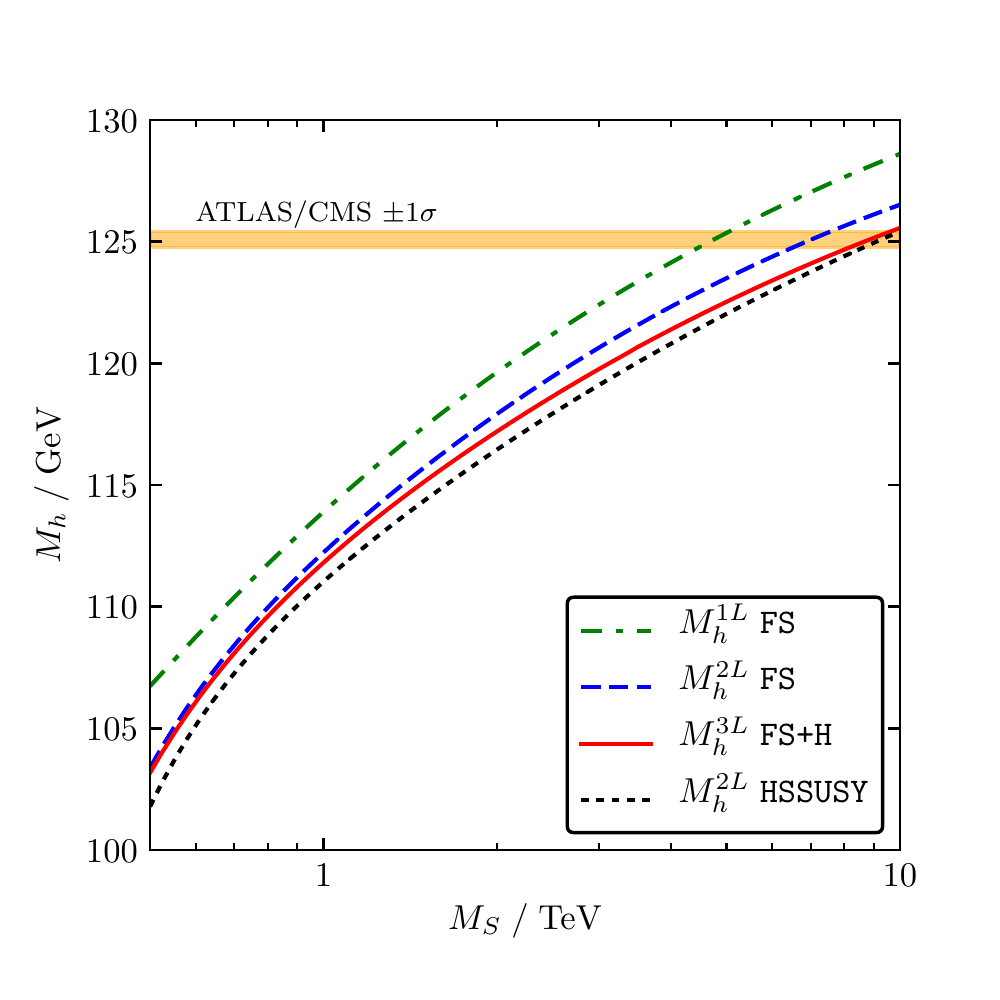}
  \caption{Comparison of the lightest Higgs pole mass calculated at
    the \one-, \two- and \three-loop level with \fs, \fsh, \htm and
    \HSSUSY as a function of the \SUSY scale for the ``heavy
    sfermions'' scenario of \citere{Kunz:2014gya}.  The horizontal
    orange band shows the measured Higgs mass
    $M_h = (125.09 \pm 0.32)\unit{GeV}$ including its experimental
    uncertainty.}
  \label{fig:Mh_heavy_sfermions}
\end{figure}

In \figref{fig:Mh_heavy_sfermions} we compare our results to the
\three-loop calculation presented in \citere{Kunz:2014gya}, assuming the
input parameters for the ``heavy sfermions'' scenario defined in detail
in the \texttt{example} folder of \citere{h3murl}.  In the left panel
the blue circles show the \htm result, including only the terms of
$\order{\at + \at\as + \at\as^2}$, where the \MSSM \DRbar top mass is
calculated using the ``running and decoupling'' procedure described in
\citere{Kunz:2014gya}.  The black crosses show the same result, except
that the \DRbar top mass at the \SUSY scale is taken from the spectrum
generator \fsh.  We can reproduce the latter result with \fsh if we take
the same terms into account, i.e., $\order{\at + \at\as + \at\as^2}$;
see the dotted red line in \figref{fig:Mh_heavy_sfermions}. The small
differences between the two results are due to the fact that \htm works
with on-shell electroweak parameters, while \fsh uses \DRbar
parameters.
The inclusion of all \one-loop contributions to $M_h$ and the momentum
iteration reduces the Higgs mass by $4$--$6\unit{GeV}$, as shown by the
red dashed line.  Including all \two- and \three-loop corrections which
are available in \fsh, i.e., $\order{(\at+\ab)\as + (\at + \ab)^2 + \atau^2 +
  (\at+\ab)\as^2}$, further reduces the Higgs mass by up to $2\unit{GeV}$, as
shown by the red solid line.\footnote{By default all available \two- and
  \three-loop corrections are included in \fsh.}
The right panel of \figref{fig:Mh_heavy_sfermions} shows again our
\one-, \two-, and \three-loop predictions obtained with \fs, \fsh, as
well as the \EFT result of \HSSUSY.  Similar to \figref{fig:Mh_MS}, we
observe that the higher-order terms lower the predicted Higgs mass and
bring it closer to the resummed result.
A detailed comparison of \fsh to a result where \htm is combined with
the lower-order results of \FeynHiggs is beyond the scope of this paper
and left to a future publication.

\begin{figure}[tbh]
  \centering \includegraphics[width=0.49\textwidth]{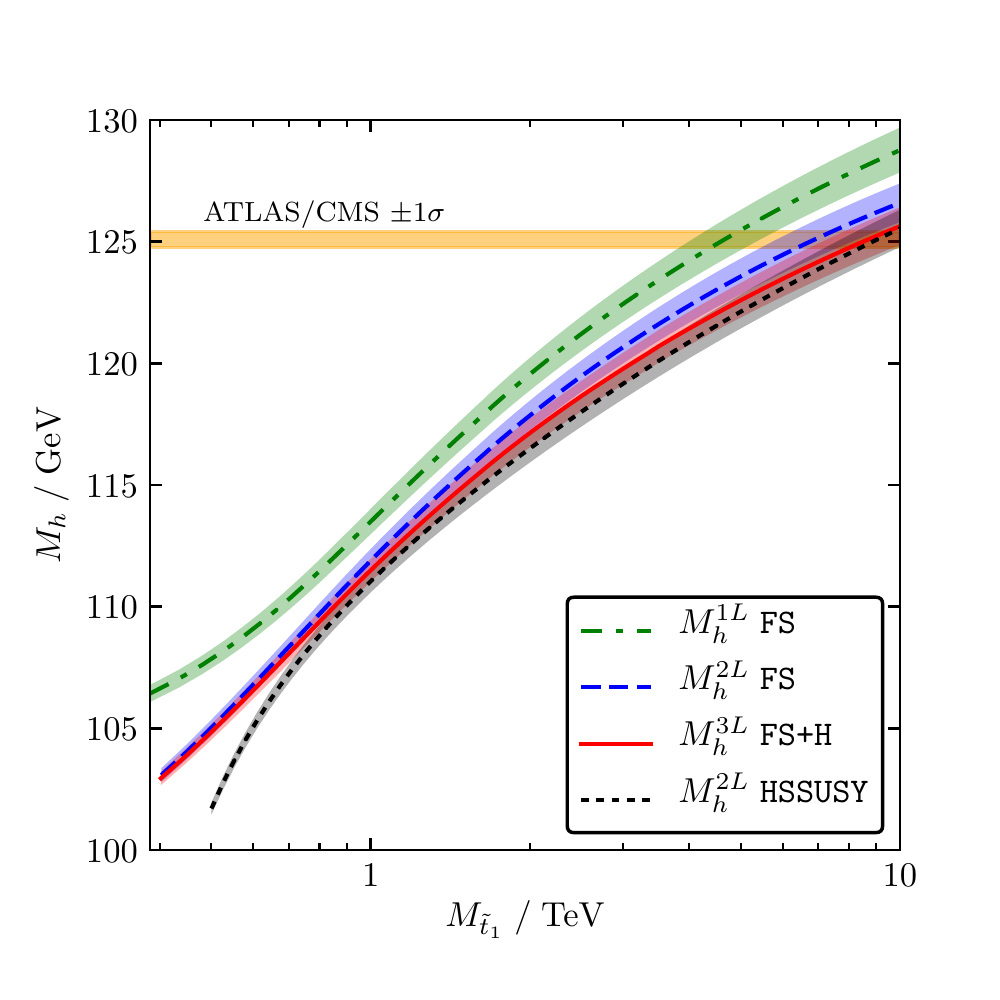}
  \caption{Comparison of the lightest Higgs pole mass calculated at
    the \one-, \two- and \three-loop level with \fs, \fsh and \HSSUSY as a
    function of the lightest stop pole mass for the benchmark point of
    Fig.\,1 of \citere{Feng:2013ena}.  The horizontal orange band shows
    the measured Higgs mass $M_h = (125.09 \pm 0.32)\unit{GeV}$
    including its experimental uncertainty.  The bands around the
    calculated Higgs mass values show the parametric uncertainty from
    $M_t = (173.34 \pm 0.98)\unit{GeV}$ and
    $\as^{\SM(5)}(M_Z) = 0.1184 \pm 0.0006$.}
  \label{fig:Mh_Mstop}
\end{figure}

\figref{fig:Mh_Mstop} shows the lightest \MSSM\ Higgs mass as obtained
by \fs\ at \one- and \two-loop level, the \fsh result, as well as the
\EFT\ prediction obtained with \HSSUSY. The \MSSM\ parameters are
defined in the \DRbar scheme and are chosen in the style of
\citere{Feng:2013ena}:\footnote{The scenario of \citere{Feng:2013ena}
  appears to be not fully defined; in particular, $M_A$ and the sfermion
  mixing parameters other than $X_t$ remain unspecified.}  The
soft-breaking mass parameters of the left- and right-handed stops are
set equal at the \SUSY scale $\MS = \sqrt{\mstop{1}\mstop{2}}$,
i.e.\ $m_{\tilde{t}_L}(\MS) = m_{\tilde{t}_R}(\MS)$. All other
soft-breaking sfermion mass parameters are set to
$\drbarmass_{\tilde{f}}(\MS) = m_{\tilde{t}_{L,R}}(\MS) + 1\unit{TeV}$.
Stop mixing is disabled, $X_t(\MS) = 0$, and the remaining
trilinear couplings are set to zero at the scale \MS.
The gaugino mass parameters, the super-potential
$\mu$ parameter and the \abbrev{CP}-odd \DRbar Higgs mass are set to
$M_1(\MS) = M_2(\MS) = M_3(\MS) = 1.5\unit{TeV}$, $\mu(\MS) =
200\unit{GeV}$ and $m_A(\MS) = \MS$, respectively, and we fix
$\tan\beta(M_Z) = 20$.  As opposed to the results shown in Fig.\,1 of
\citere{Feng:2013ena},\footnote{Note that, in contrast to
  \citere{Feng:2013ena}, we are using a logarithmic scale in
  \figref{fig:Mh_Mstop}.} we observe a
reduction of $M_h$ towards higher loop orders, thus leading to
the opposite conclusion of a heavy \SUSY\ spectrum in this
scenario, given the current experimental value for the Higgs
mass. Reassuringly, the higher order corrections move the fixed-order
result closer to the resummed result, leading to agreement between
the two at the level of about $1\unit{GeV}$ even at comparatively large
\SUSY scales.

\section{Conclusions}\label{sec:conclusions}

We have presented the implementation \himalaya of the \three-loop
  $\order{\atasas + \abasas}$ terms
  of \citeres{Harlander:2008ju,Kant:2010tf} for the light
\abbrev{CP}-even Higgs mass in the \MSSM, and its combination with
  the \DRbar spectrum generator framework \fs.  These \three-loop
contributions have been available in the public program \htm before,
where they were combined with the on-shell calculation of \FeynHiggs.
With the implementation into \fs presented here, we were able to study
the size of the
\three-loop contributions within a pure \DRbar environment.  Despite the
fact that the genuine $\order{\at\as^2}$ corrections are
positive\,\cite{Kant:2010tf}, the combination with the \two-loop
decoupling terms in the top Yukawa coupling lead to an overall reduction
of the Higgs mass prediction relative to the ``original'' \two-loop \fs
result by about 2\,GeV, depending on the value of the stop masses and
the stop mixing. This moves the fixed-order prediction for the Higgs
mass significantly closer to the result obtained from a
pure \EFT calculation in the region where both approaches are
expected to give sensible results.  Contributions of order
$\order{\abasas}$ are found to be negligible in all scenarios studied
here.

To indicate the remaining theory uncertainty due to higher order
effects, we have varied the renormalization scale which enters the
calculation by a factor \two.  The results show that the inclusion of
the \three-loop contributions reduces the scale uncertainty of the Higgs
mass by around a factor \two, compared to a calculation without the
genuine \three-loop effects.  We conclude that our implementation leads
to an improved \abbrev{CP}-even Higgs mass prediction relative to
the \two-loop results. Our implementation of the \three-loop terms
should be useful also for other groups that aim at a high-precision
determination of the Higgs mass in \SUSY models.

\section*{Acknowledgments}

We would like to thank Luminita Mihaila, Matthias Steinhauser, and
Nikolai Zerf for helpful comments on the manuscript, and valuable help
in the comparison with \htm. Further thanks go to Pietro Slavich for his
valuable comments, in particular for pointing out an inconsistency in
Section\,\ref{sec:determination_of_MSSM_parameters} of the original
manuscript. Alexander Bednyakov kindly provided the general
\two-loop \SQCD corrections to the running top and bottom masses in the
\MSSM in \mathematica format.  \abbrev{RVH} would like to thank the
theory group at \abbrev{NIKHEF}, where part of this work was done, for
their kind hospitality.  \abbrev{AV} would like to thank the Institute
for Theoretical Physics (\abbrev{ITP}) in Heidelberg for its warm
hospitality.  Financial support for this work was provided by
\abbrev{DFG}.

\appendix

\section{Installation of \himalaya}

\himalaya can be downloaded as compressed package from
\cite{himalaya}.  After the package has been extracted, \himalaya can
be configured and compiled by running
\begin{lstlisting}
cd $HIMALAY_PATH
mkdir build
cd build
cmake ..
make
\end{lstlisting}
where \code{$HIMALAY_PATH} is the path to the \himalaya directory.
When the compilation has finished, the build directory will contain the
\himalaya library \code{libHimalaya.a}.  For convenience, a library
named \code{libDSZ.a} is created in addition, which contains the
\two-loop $\order{\at\as}$ corrections from \citere{Degrassi:2001yf}.

\section{Installation of \fs with \himalaya}

We provide a dedicated version of \fs 1.7.4, which uses
\himalaya to calculate the Higgs pole mass at the \three-loop
level.  This package contains three pre-generated \MSSM
models:
\begin{itemize}
\item \texttt{MSSMNoFVHimalaya}: This
  model represents the \MSSM without (s)fermion flavour violation, where
  $\tan\beta$ is fixed at the scale $M_Z$ and the other \SUSY parameters
  are fixed at a user-defined input scale.  The parameters $\mu$ and
  $B\mu$ are fixed by the electroweak symmetry breaking conditions.  The
  \SUSY mass spectrum, including the Higgs pole masses, is calculated at
  the scale $Q = \sqrt{\mstop{1} \mstop{2}}$, where $\mstop{i}$ are the
  two $\DRbar$ stop masses.
\item \texttt{MSSMNoFVatMGUTHimalaya}: This is the same model as the
  \texttt{MSSMNoFVHimalaya}, except that the input scale is the \abbrev{GUT} scale
  $M_X$, defined to be the scale where $g_1(M_X) = g_2(M_X)$.
\item \texttt{NUHMSSMNoFVHimalaya}: This is the same model as the
  \texttt{MSSMNoFVHimalaya}, except that the soft-breaking Higgs mass
  parameters $m_{H_u}^2$ and $m_{H_d}^2$ are fixed by the electroweak
  symmetry breaking conditions.
\end{itemize}
The package \texttt{FlexibleSUSY-1.7.4-Himalaya.tar.gz} can be
downloaded from \citere{flexiblesusy}.
To extract the package at the command line, run
\begin{lstlisting}
tar -xf FlexibleSUSY-1.7.4-Himalaya.tar.gz
cd FlexibleSUSY-1.7.4-Himalaya/
\end{lstlisting}
After the extraction, \fs must be configured and compiled by running
\begin{lstlisting}
./configure \
   --with-himalaya-incdir=$HIMALAY_PATH/source/include \
   --with-himalaya-libdir=$HIMALAY_PATH/build
make
\end{lstlisting}
See \code{./configure --help} for more options.  One can use \code{make
  -j<N>} to speed-up the compilation if \code{<N>} \abbrev{CPU} cores
are available.  When the compilation has finished, the \MSSM spectrum
generators can be run from the command line as
\begin{lstlisting}
models/MSSMNoFVHimalaya/run_MSSMNoFVHimalaya.x \
  --slha-input-file=models/MSSMNoFVHimalaya/LesHouches.in.MSSMNoFVHimalaya
  --slha-output-file=LesHouches.out.MSSMNoFVHimalaya
\end{lstlisting}
The file \texttt{LesHouches.out.MSSMNoFVHimalaya} will then contain the
\SUSY particle spectrum in \abbrev{SLHA} format.  Alternatively, the
\mathematica interface of \fs can be used:
\begin{lstlisting}
math -run "<< \"models/MSSMNoFVHimalaya/run_MSSMNoFVHimalaya.m\""
\end{lstlisting}
For each model an example \abbrev{SLHA} input file and an example \mathematica
script can be found in \texttt{models/<model>/}.

\section{Configuration options to calculate the Higgs mass at \three-loop
  level with \fs}
\label{app:FS_options}

To calculate the \abbrev{CP}-even Higgs pole masses at order
$\order{\atasas + \abasas}$ at the scale $Q=\MS$, the top and bottom
Yukawa couplings $y_t(\MS)$ and $y_b(\MS)$ as well as the strong
coupling constant $\as(\MS)$ must be extracted from the input
parameters at the appropriate loop level.

\paragraph{Strong coupling constant.}
To calculate $M_h$ at the \three-loop level at $\order{\atasas + \abasas}$
correctly, $\as(\MS)$ must be extracted at the \one-loop level from the
input value $\as^{\SM(5)}(M_Z)$ as described in
\secref{sec:determination_of_MSSM_parameters}.  To achieve that in
\fs, the global threshold correction loop order (\code{EXTPAR[7]})
must be set to $1$ (or higher) and the specific threshold correction
loop order for $\as$ ($3$rd digit from the right in \code{EXTPAR[24]})
must be set to $1$ (or higher) in the \abbrev{SLHA} input file.  See
the next paragraph for an example.

\paragraph{Top and bottom Yukawa couplings.}
\fs by default determines $y_t(M_Z)$ from the top pole mass at the
full \one-loop level including \two-loop Standard Model \QCD corrections, see
\citere{Athron:2014yba}.  The bottom Yukawa coupling $y_b(M_Z)$ is
determined at the full \one-loop level from the running bottom quark mass
in the Standard Model with five active quark flavours,
$m_b^{\SM(5),\MSbar}(m_b)$, where $\tan\beta$-enhanced higher order
corrections are resummed.
Both calculations are not sufficient for the calculation of $M_h$ at
the \three-loop level at $\order{\atasas + \abasas}$, because strong \two-loop
corrections from \SUSY particles would be missing.  For this reason, the complete
\two-loop strong corrections to the top and bottom Yukawa couplings of
\citeres{Bednyakov:2002sf,Bednyakov:2005kt,Bednyakov:2007vm,Bauer:2008bj} have been
implemented into \fs.  They must be activated by setting the global
threshold correction loop (\code{EXTPAR[7]}) order to $2$ and by
setting the threshold correction loop order for $y_t$ and $y_b$ ($7$th
and $8$th digit from the right in \code{EXTPAR[24]}) to $2$ in the
\abbrev{SLHA} input file:
\begin{lstlisting}
Block FlexibleSUSY
    7   2         # threshold corrections loop order
   24   122111121 # individual threshold correction loop orders
\end{lstlisting}
In the \mathematica interface of \fs these two settings are controlled
using the \code{thresholdCorrectionsLoopOrder} and
\code{thresholdCorrections} symbols:
\begin{lstlisting}
handle = FS<model>OpenHandle[
    fsSettings -> {
        thresholdCorrectionsLoopOrder -> 2,
        thresholdCorrections -> 122111121
    }
    ...
];
\end{lstlisting}
Here, \texttt{<model>} is the used \fs model from above, i.e.\ either
\texttt{MSSMNoFVHimalaya}, \texttt{MSSMNoFVatMGUTHimalaya} or
\texttt{NUHMSSMNoFVHimalaya}.

\paragraph{Three-loop corrections to the \abbrev{CP}-even Higgs mass.}
To use the \three-loop corrections of order $\order{\atasas + \abasas}$
to the light \abbrev{CP}-even Higgs mass in
the \MSSM from \citeres{Harlander:2008ju,Kant:2010tf}, the pole mass and
\abbrev{EWSB} loop orders must be set to $3$ in the \abbrev{SLHA} input
file.  In addition, the individual \three-loop corrections should be
switched on, by setting the flags $26$ and $27$ to $1$.  The user can
select between the \DRbar and \MDRbar scheme for the \three-loop
corrections by setting the flag $25$ to $0$ or $1$, respectively:
\begin{lstlisting}
Block FlexibleSUSY
    4   3   # pole mass loop order
    5   3   # EWSB loop order
   25   0   # ren. scheme for Higgs 3L corrections (0 = DR, 1 = MDR)
   26   1   # Higgs 3-loop corrections O(alpha_t alpha_s^2)
   27   1   # Higgs 3-loop corrections O(alpha_b alpha_s^2)
\end{lstlisting}
In the \mathematica interface of \fs the pole mass and \abbrev{EWSB} loop
orders are controlled using the \code{poleMassLoopOrder} and
\code{ewsbLoopOrder} symbols, respectively.  The individual \three-loop
corrections can be switched on/off by using the
\code{higgs3loopCorrectionAtAsAs} and
\code{higgs3loopCorrectionAbAsAs} symbols.  The renormalization scheme
is controlled by \code{higgs3loopCorrectionRenScheme}.  The above
shown \abbrev{SLHA} input settings read in \fs's \mathematica interface
\begin{lstlisting}
handle = FS<model>OpenHandle[
    fsSettings -> {
        poleMassLoopOrder -> 3,
        ewsbLoopOrder -> 3,
        higgs3loopCorrectionRenScheme -> 0,
        higgs3loopCorrectionAtAsAs -> 1,
        higgs3loopCorrectionAbAsAs -> 1
    }
    ...
];
\end{lstlisting}

\paragraph{Three-loop renormalization group equations.}
Optionally, the known \three-loop renormalization group equations can be
used to evolve the \MSSM \DRbar parameters from $M_Z$ to $\MS$
\cite{Jack:2003sx,Jack:2004ch}.  To activate the \three-loop \abbrev{RGEs}, the
$\beta$ function loop order must be set to $3$ in the \abbrev{SLHA} input file:
\begin{lstlisting}
Block FlexibleSUSY
    6   3   # beta-functions loop order
\end{lstlisting}
In the \mathematica interface of \fs the $\beta$ function loop order
is controlled using the \code{betaFunctionLoopOrder} symbol:
\begin{lstlisting}
handle = FS<model>OpenHandle[
    fsSettings -> {
        betaFunctionLoopOrder -> 3
    }
    ...
];
\end{lstlisting}

\paragraph{Recommended configuration options for \fsh.}
We recommend to run \fsh with the following SLHA configuration
options:
\begin{lstlisting}
Block FlexibleSUSY
    4   3         # pole mass loop order
    5   3         # EWSB loop order
    6   3         # beta-functions loop order
    7   2         # threshold corrections loop order
   24   122111121 # individual threshold correction loop orders
   25   0         # ren. scheme for 3L corrections (0 = DR, 1 = MDR)
   26   1         # Higgs 3-loop corrections O(alpha_t alpha_s^2)
   27   1         # Higgs 3-loop corrections O(alpha_b alpha_s^2)
\end{lstlisting}
At the \mathematica level we recommend to use:
\begin{lstlisting}
handle = FS<model>OpenHandle[
    fsSettings -> {
        poleMassLoopOrder -> 3,
        ewsbLoopOrder -> 3,
        betaFunctionLoopOrder -> 3,
        thresholdCorrectionsLoopOrder -> 2,
        thresholdCorrections -> 122111121,
        higgs3loopCorrectionRenScheme -> 0,
        higgs3loopCorrectionAtAsAs -> 1,
        higgs3loopCorrectionAbAsAs -> 1
    }
    ...
];
\end{lstlisting}

\section{\himalaya interface}

\paragraph{Input parameters.}
To calculate the three-loop corrections to the light \abbrev{CP}-even
Higgs pole mass at order $\order{\atasas + \abasas}$ with \himalaya,
the set of \DRbar parameters is needed, which is shown in the
following code snippet. The parameters are stored in the \code{struct}
\code{Parameters} which contains the following members:
\begin{lstlisting}[language=C++]
typedef Eigen::Matrix<double,2,1> V2;
typedef Eigen::Matrix<double,2,2> RM22;
typedef Eigen::Matrix<double,3,3> RM33;

struct Parameters {
   // DR-bar parameters
   double scale{};         // renormalization scale
   double mu{};            // mu parameter
   double g3{};            // gauge coupling g3 SU(3)
   double vd{};            // VEV of down Higgs
   double vu{};            // VEV of up Higgs
   RM33 mq2{RM33::Zero()}; // soft-breaking squared left-handed squark
                           // mass parameters
   RM33 md2{RM33::Zero()}; // soft-breaking squared right-handed
                           // down-squark mass parameters
   RM33 mu2{RM33::Zero()}; // soft-breaking squared right-handed
                           // up-squark mass parameters
   double At{};            // trilinear stop-Higgs coupling
   double Ab{};            // trilinear sbottom-Higgs coupling

   // DR-bar masses
   double MG{};            // gluino
   double MW{};            // W
   double MZ{};            // Z
   double Mt{};            // top quark
   double Mb{};            // down quark
   double MA{};            // CP-odd Higgs
   V2 MSt{nan, nan};       // stops
   V2 MSb{nan, nan};       // sbottoms

   // DR-bar mixing angles
   double s2t{nan};        // sine of 2 times the stop mixing angle
   double s2b{nan};        // sine of 2 times the sbottom mixing angle
};
\end{lstlisting}
All these parameters are given at the scale stored in the \code{scale}
variable, which is typically the \SUSY scale. The input values of the
stop/sbottom masses and their associated mixing angle are optional, so
their default value is set to \code{nan}
(\code{std::numeric_limits<T>::quiet_NaN()}). If no input is provided,
the \DRbar stop masses will be calculated by diagonalizing the stop
mass matrix
\begin{equation}
  \mathcal{M}_{\tilde{t}} =
  \begin{pmatrix}
    \big(m_{\tilde{Q}}^2\big)_{33} + m_t^2 + g_{t}M_Z^2c_{2\beta} & \tilde{X}_{t} \\
    \tilde{X}_{t} & \big(m_{\tilde{u}}^2\big)_{33} + m_t^2 + Q_ts_W^2M_Z^2c_{2\beta}
  \end{pmatrix}.
  \label{eq:stop_mass_matrix}
\end{equation}
Here, $(m_{\tilde{Q}})_{33}$ is the left third generation scalar quark mass parameter,
$g_{t} = 1/2 - Q_ts_W^2$, $\tilde{X}_t = m_t(A_t - \mu\cot\beta)$,
$(m_{\tilde{u}})_{33}$ the right scalar top mass parameter, $Q_t = 2/3$, $s_W$
the sine of the weak mixing angle and
$c_{2\beta}=\cos(2\beta)$.  The sbottom mass matrix is obtained by
replacing $t\to b$ and $\tilde{u}\to\tilde{d}$ in
\eqref{eq:stop_mass_matrix} with $g_b=-(1/2+Q_bs_W^2)$,
$\tilde{X}_b = m_b(A_b - \mu \tan\beta)$ and $Q_b = -1/3$.

\paragraph{Calculation of the three-loop corrections.}
All the functions which are required for the calculation of the
three-loop corrections are implemented as methods of the class
\code{HierarchyCalculator}.

In the context of \himalaya, the procedure described in
\secref{sec:higgsMass3L} is implemented by the member function
\begin{lstlisting}[language=C++]
HierarchyObject ho = HierarchyCalculator::calculateDMh3L(bool isAlphab,
   int mdrFlag);
\end{lstlisting}
Here, the integer \code{mdrFlag} is optional and can be used to switch between the
$\DRbar$- (0) and the $\MDRbar$-scheme (1). The $\DRbar$-scheme is chosen as default.  The returned object holds
all information of the hierarchy selection process, such as the best
fitting hierarchy, or the relative error
$\delta_{i_0}^{\mathrm{2L}}/M^{\mathrm{DSZ}}_{h}$, where
$\delta_i^\mathrm{2L}$ is defined in \eqn{eq:dsz_error}, and $i_0$
denotes the ``optimal'' hierarchy as determined by the procedure of
\secref{sec:hierarchy}. The latter represents a lower limit on the
expected accuracy of the expansion by comparison to the exact \two-loop
result $M_h^\mathrm{DSZ}$. In addition to that, the
\code{HierarchyObject} offers a set of member functions which provide
access to all intermediate results. These functions are summarized in
\tabref{tab:hierarchyObjectMemberFunctions}.
\begin{table}[h]
  \begin{center}
    \caption[]{\label{tab:hierarchyObjectMemberFunctions} Description
      of the member functions of the \code{HierarchyObject} class.}
    {\renewcommand{\arraystretch}{1.3}%
    \begin{tabularx}{\textwidth}{lX}
      \toprule
      Function name & Returned value \\
      \midrule
      \code{getIsAlphab()} 	&Returns the \code{bool} \code{isAlphab}.\\
      \code{getSuitableHierarchy()}	&Returns the suitable hierarchy as an \code{int}.\\
      \code{getAbsDiff2L()}	&Returns the \code{double}
    $\delta_{i_0}^{\mathrm{2L}}$ for the suitable hierarchy.\\
      \code{getRelDiff2L()}	&Returns the \code{double}
    $\delta_{i_0}^{\mathrm{2L}}/M^{\mathrm{DSZ}}_{h}$ for the suitable
    hierarchy.\\
      \code{getExpUncertainty(int loops)}	&Returns the uncertainty of the expansion at the given 
      loop order (cf. \secref{sec:hierarchy}).\\
      \code{getDMh(int loops)}	&Returns the Higgs mass matrix proportional to $\at$ or $\ab$
      at the given loop order. Note that at the \two-loop level only corrections of order $\mathcal{O}(\at\as)$ are considered.\\

      \code{getDRToMDRShift()} & Returns the loop correction to the
      Higgs mass matrix to convert from the $\DRbar$ to $\MDRbar$
      scheme, according to \eqn{eq:DRToMDRShift}. The $\MDRbar$
      corrections are of order
      $\order{\as + \as^2}$ by convention.\\

      \code{getMDRMasses()}	&Returns the vector of $\MDRbar$ masses
      $\{\mstopmod{1}, \mstopmod{2}\}$ ($\{\msbotmod{1}, \msbotmod{2}\}$),
      if \code{isAlphab} is \code{false} (\code{true}).\\
      \bottomrule
    \end{tabularx}}
  \end{center}
\end{table}
The selection method described in \secref{sec:higgsMass3L} is also
applied to the (s)bottom
contributions by replacing $t \rightarrow b$, so that only terms of
order $\oabas$ are considered in the comparison. By setting the Boolean
parameter \code{isAlphab} to \code{false} (\code{true}) the
\code{calculateDMh3L} function returns the \code{HierarchyObject} for
the loop corrections proportional to $\at$ ($\ab$).

\emph{Example:} Function calls for the benchmark point \abbrev{SPS2}:
\begin{lstlisting}[language=C++]
#include "HierarchyCalculator.hpp"
#include "HierarchyObject.hpp"

h3m::Parameters setupSPS2()
{
   h3m::Parameters pars;

   pars.scale = 1.11090135E+03;
   pars.mu = 3.73337018E+02;
   pars.g3 = 1.06187116E+00;
   pars.vd = 2.51008404E+01;
   pars.vu = 2.41869332E+02;
   pars.mq2 << 2.36646981E+06, 0, 0,
               0, 2.36644973E+06, 0,
               0, 0, 1.63230152E+06;
   pars.md2 << 2.35612778E+06, 0, 0,
               0, 2.35610884E+06, 0,
               0, 0, 2.31917415E+06;
   pars.mu2 << 2.35685097E+06, 0, 0,
               0, 2.35682945E+06, 0,
               0, 0, 9.05923409E+05;
   pars.Ab = -784.3356416708631;
   pars.At = -527.8746242245387;

   pars.MA = 1.48446235E+03;
   pars.MG = 6.69045022E+02;
   pars.MW = 8.04001915E+01;
   pars.MZ = 8.97608307E+01;
   pars.Mt = 1.47685846E+02;
   pars.Mb = 2.38918959E+00;
   pars.MSt << 9.57566721E+02, 1.28878643E+03;
   pars.MSb << 1.27884964E+03, 1.52314587E+03;
   pars.s2t = sin(2*asin(1.13197339E-01));
   pars.s2b = sin(2*asin(-9.99883015E-01));

   return pars;
}

int main() {
   h3m::HierarchyCalculator hc(setupSPS2());

   // get the HierarchyObject with entries proportional to alpha_t
   // in the DR scheme
   auto hoTop = hc.calculateDMh3L(false);

   // get the 3-loop correction O(alpha_t * alpha_s^2)
   auto DMh_top_3L = hoTop.getDMh(3);
}
\end{lstlisting}

\paragraph{Estimation of the uncertainty of the expansion.}
\label{sec:expUncertainty}
In addition to the relative error of the hierarchy choice
$\delta_{i_0}^\mathrm{2L}/M_h^\text{DSZ}$ (see above), we provide a
member function which returns a measure for the quality of convergence
of the expansion at a given loop order, given by
$\delta^\text{conv}_{i_0}$ defined in \eqn{eq:exp_error}, where again
$i_0$ labels the ``optimal'' hierarchy. It can be called with
\begin{lstlisting}[language=C++]
Eigen::Matrix2d HierarchyCalculator::getExpansionUncertainty(
   HierarchyObject ho, const Eigen::Matrix2d& massMatrix
   int oneLoopFlag, int twoLoopFlag, int threeLoopFlag);
\end{lstlisting}
Its arguments are a \code{HierarchyObject}, the Higgs mass matrix
\code{massMatrix} up to the loop order of interest, and three flags
(\code{oneLoopFlag}, \code{twoLoopFlag}, \code{threeLoopFlag}) to define
the desired loop orders. Using the member function \code{calculateDMh},
the returned \code{HierarchyObject} provides the user with the quantity
$\delta_{i_0}^\text{conv}$ at \two\ and \three\ loops by
default.

\emph{Example:} For the benchmark point \abbrev{SPS2} one could estimate the
uncertainty by calling
\begin{lstlisting}[language=C++]
...
// get the HierarchyObject with entries proportional to alpha_t
// in the DR scheme
auto hoTop = hc.calculateDMh3L(false);

// get the expansion uncertainty for the
// 3-loop correction O(alpha_t * alpha_s^2)
auto expansionUncertainty3LTop = hoTop.getExpUncertainty(3);

// calculate the expansion uncertainty for the
// 1-loop correction O(alpha_t)
auto expansionUncertainty1LTop = hc.getExpansionUncertainty(hoTop,
   ho.getDMh(0), 1, 0, 0);
\end{lstlisting}

\bibliographystyle{JHEP}
\bibliography{paper}

\end{document}